\documentclass[reprint,superscriptaddress,amsmath,amssymb,aps,prl,floatfix]{revtex4-2}
\usepackage{float}
\usepackage{graphicx}
\usepackage[svgnames]{xcolor}
\usepackage[bookmarks=false]{hyperref}
\usepackage{dcolumn}
\usepackage{bm}
\graphicspath{{Figures/}}

\usepackage{newfloat}
\DeclareFloatingEnvironment[name={Supplementary Figure},fileext=lof]{suppfigure}
\DeclareFloatingEnvironment[name={Supplementary Table},fileext=lof]{supptable}

\begin{document}
\title{Thermal resistivity and hydrodynamics of the degenerate electron fluid in antimony}

\author{Alexandre Jaoui}
\email{alexandre.jaoui@espci.fr}
\affiliation{JEIP, USR 3573 CNRS, Coll\`ege de France, PSL Research University, 11, Place Marcelin Berthelot, 75231 Paris Cedex 05, France}
\affiliation{Laboratoire de Physique et Etude des Mat\'eriaux (CNRS/UPMC), Ecole Sup\'erieure de Physique et de Chimie Industrielles, 10 Rue Vauquelin, 75005 Paris, France}
\author{Beno\^it Fauqu\'e}
\affiliation{JEIP, USR 3573 CNRS, Coll\`ege de France,
PSL Research University, 11, Place Marcelin Berthelot,
75231 Paris Cedex 05, France}
\author{Kamran Behnia}
\affiliation{Laboratoire de Physique et Etude des Mat\'eriaux (CNRS/UPMC), Ecole Sup\'erieure de Physique et de Chimie Industrielles, 10 Rue Vauquelin, 75005 Paris, France}
\date{\today}

\begin{abstract}
Detecting hydrodynamic fingerprints in the flow of electrons in solids constitutes a dynamic field of investigation in contemporary condensed matter physics. Most attention has been focused on the regime near the degeneracy temperature when the thermal velocity can present a spatially modulated profile. Here, we report on the observation of a hydrodynamic feature in the flow of quasi-ballistic degenerate electrons in bulk antimony. By scrutinizing the temperature dependence of thermal and electric resistivities, we detect a size-dependent departure from the Wiedemann-Franz law,  unexpected in the momentum-relaxing picture of transport. This observation finds a natural explanation in the hydrodynamic picture, where upon warming, momentum-conserving collisions reduce quadratically in temperature both viscosity and thermal diffusivity. This effect has been established theoretically and experimentally in normal-state liquid $^3$He. The comparison of electrons in antimony and fermions in $^3$He paves the way to a quantification of momentum-conserving fermion-fermion collision rate in different Fermi liquids.
\end{abstract}
\maketitle

\section{Introduction}

The possibility of viscous electronic flow, suggested long ago by Gurzhi~\cite{gurzhi1968}, has attracted a lot of attention recently ~\cite{hartnoll2015,principi2015,scaffidi2017}. When momentum-conserving (MC) collisions among electrons outweigh scattering by boundaries as well as various momentum-relaxing (MR) collisions, the quasiparticle (QP) flow profile is expected to change. In this case, momentum and energy of the QPs will be redistributed over a length much shorter than the resistive mean free path. As a consequence, the further away the electron is from the boundaries, the hardest the MC collisions will make it for the QP to make its way to the boundaries of the system. If boundary scattering becomes also more frequent than MR collisions, then the QPs the furthest away from the boundaries are less likely to undergo a dissipative collision. As a consequence, the QP flow becomes analogous to that of a viscous fluid in a channel (dubbed the Poiseuille flow). Such viscous corrections to electronic transport properties have been seen by a number of experiments~\cite{molenkamp1994,moll2016,crossno2016,bandurin2016,gooth2018,sulpizio2019}. All these studies were performed on mesoscopic ultra-pure metals. The strongest hydrodynamic signatures have been seen in graphene near the neutrality point and when electron velocity is set by the thermal energy. The velocity of degenerate electrons, on the other hand, is narrowly distributed around the Fermi velocity. Moreover, since the rate of electron-electron collisions is proportional to the square of the ratio of temperature to the Fermi temperature, MR collisions rarefy with increasing degeneracy.

Nevertheless, quantum liquids (such as both isotopes of helium) present hydrodynamic features associated with viscosity. Soon after the conception of Landau's Fermi liquid theory, Abrikosov and Khalatnikov~\cite{Abrikosov_1959} calculated the transport coefficients of an isotropic Fermi liquid, focusing on liquid $^3$He well below its degeneracy temperature. They showed that since the phase space for fermion-fermion scattering grows quadratically with temperature $T$, viscosity $\eta$ (which is the diffusion constant for momentum) and thermal diffusivity $D$ (which is the diffusion constant for energy) both follow $T^{-2}$ and, as result, $\kappa \propto T^{-1}$. Subsequent theoretical studies~\cite{Nozieres,Brooker1968} confirmed this pioneering study and corrected~\cite{Brooker1968} the prefactors. Thermal conductivity~\cite{Wheatley1968,greywall1984} and viscosity~\cite{bertinat1974,alvesalo1975} measurements at very low temperatures found the theoretically predicted temperature dependence for both quantities below $T=0.1$K, deep inside the degenerate regime.

\begin{figure*}
\centering
\makebox{\includegraphics[width=1\textwidth]{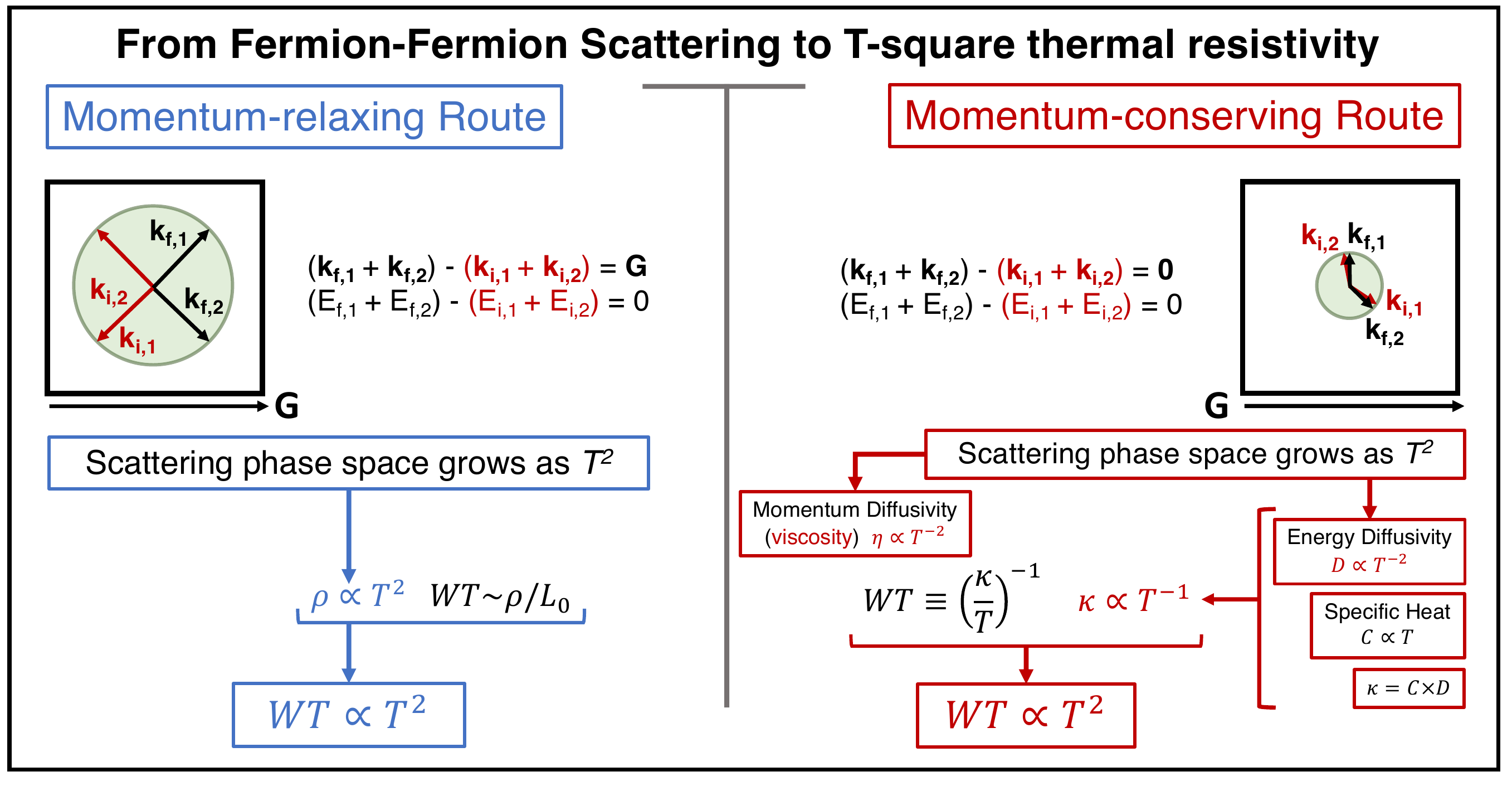}}
    \caption{\textbf{Two routes towards $T^2$ thermal resistivity.} $T$-square thermal resistivity in a Fermi liquid can arise in two distinct pictures of transport. The momentum-relaxing picture (left) is the one commonly used in metals. Because of the presence of a lattice, Umklapp collisions among electrons can occur. \textbf{k}$\bm{_{i,j}}$ and $E_{i,j}$ respectively refer to the initial momentum and energy of electron $j$ while \textbf{k}$\bm{_{f,j}}$ and $E_{f,j}$ correspond to its final momentum and energy. These collisions decay the momentum current because a unit vector of the reciprocal lattice $\textbf{G}$ is lost during the collision. The momentum-conserving picture (right) has been applied to the fermionic quasiparticles in $^3$He. We argue that it becomes relevant to metals, provided that : i) Umklapp collisions become rare or impossible (because of the smallness of the Fermi radius) and ii) the mean free path approaches the sample size.}
    \label{Fig_T_sq}
\end{figure*}

However, the common picture of transport in metallic solids does not invoke viscosity (Fig.\ref{Fig_T_sq}). The phase space for collisions among electronic quasiparticles is also proportional to the square of temperature. But the presence of a crystal lattice alters the context. Electron-electron collisions can degrade the flow of charge and heat by transferring momentum to the underlying crystal, if there is a finite amount of disorder. We will see below that if the electronic mean free path is sufficiently long compared to the sample dimensions, and if a significant portion of collisions conserve momentum (by avoiding Umklapp processes), then a finite $\kappa T|_0$, equivalent to quadratic thermal resistivity ($WT=(\frac{\kappa}{T})^{-1}$), caused by momentum-conserving collisions and evolving hand-in-hand with viscosity becomes relevant.

A fundamental correlation between the electronic thermal conductivity $\kappa_{e}$ and the electrical conductivity $\sigma$ is given by the Wiedemann-Franz (WF) law:
\begin{equation}\label{WF}
\frac {\kappa_{e}} {\sigma T}= \frac {\pi^2} {3}\frac {k_B^2} {e^2}
\end{equation}

The left hand of the equation is the (electronic) Lorenz number, $L_e$, which can be measured experimentally. The right hand side is a fundamental constant, called the Sommerfeld value $L_0=2.44 \times 10^{-8}$V$^2$.K$^{-2}$. The WF law is expected to be valid when inelastic scattering is absent, i.e. at zero temperature.

Principi and Vignale (PV)~\cite{principi2015} recently argued that in hydrodynamic electron liquids, the WF law is violated because MC electron-electron ($e-e$) scattering would degrade thermal current but not electrical current. As a consequence, by drastically reducing the $L_e/L_0$ ratio, electron hydrodynamics would lead to a finite-temperature departure from the WF law. However, the standard transport picture based on MR collisions expects a similar departure at finite temperature as a consequence of inelastic small-angle $e-e$ scattering~\cite{ziman1972,wagner1971, paglione2005,jaoui2018,li2018}. The two pictures differ in an important feature: the evolution of the $L_e/L_0$ ratio with the carrier lifetime. In the hydrodynamic picture, the deviation from the WF law becomes more pronounced with the relative abundance of MC $e-e$ collisions, which can be amplified by reducing the weight of MR collisions (by enhancing purity or size).

Here, we present a study of heat and charge transport in semi-metallic antimony (Sb) and find that $\kappa$ and $\sigma$ both increase with sample size. Sb is the most magnetoresistant semi-metal~\cite{fauque2018}. The mean-free-path $\ell_0$ of its extremely mobile charge carriers depends on the thickness of the sample at low temperature~\cite{Bogod1973}. We begin by verifying the validity of the WF law in the zero-temperature limit and resolving a clear departure from it at finite temperature. This arises because of the inequality between the prefactors of the $T$-square electrical and thermal resistivities~\cite{jaoui2018}. In contrast to its electrical counterpart, the $T$-square thermal resistivity (which is equivalent to $\kappa \propto T^{-1}$), can be purely generated by MC scattering which sets the viscosity of the electronic liquid. We find that the departure from the WF law is amplified with the increase in the sample size and the carrier mean free path, in agreement with the hydrodynamic scenario~\cite{principi2015}. We then quantify $\kappa T|_0$ and the quadratic lifetime of fermion-fermion collisions, $\tau_{\kappa}T^2$, for electrons in Sb and compare it with that of $^3$He fermions.

\begin{figure*}
\centering
\makebox{\includegraphics[width=0.9\textwidth]{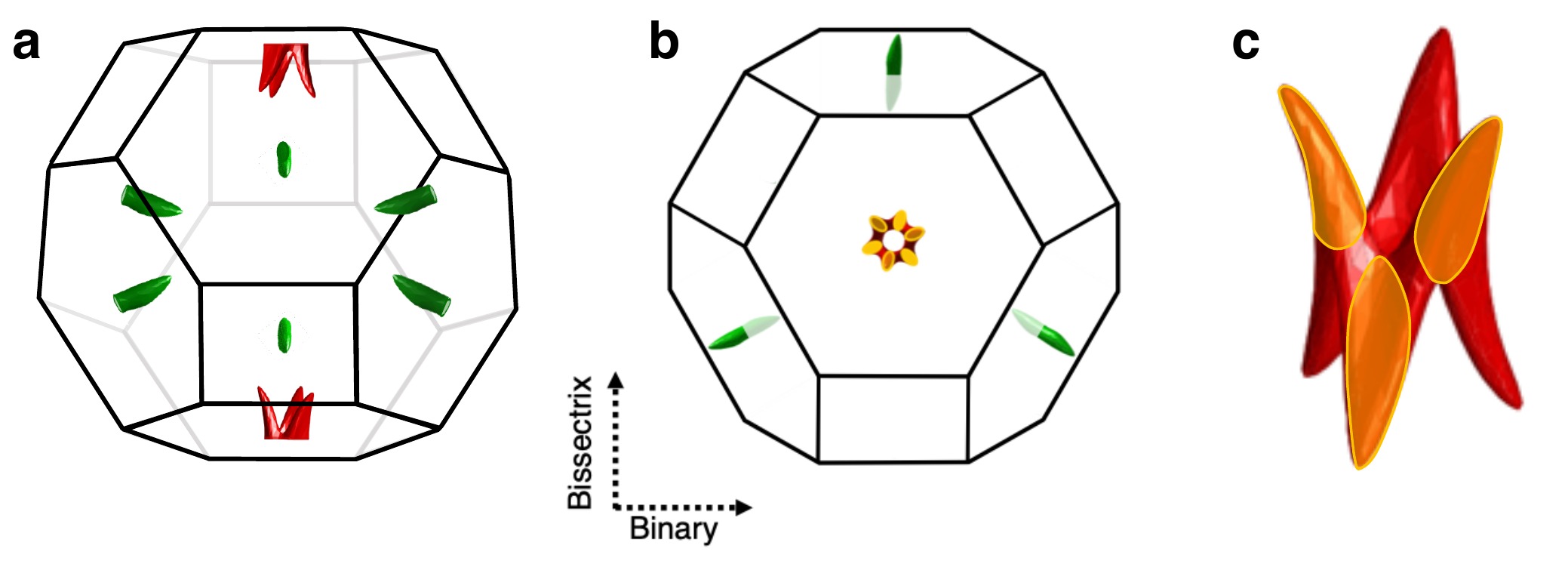}}
    \caption{\textbf{Fermi surface and the Brillouin zone of antimony (Sb).} \textbf{a}~The Fermi surface consists of electron pockets (in green) and hole pockets (in red). All pockets are located at zone boundaries and have a mirror counterpart due to the inversion symmetry. The Brillouin zone of the A7 crystal structure, nearly a truncated cuboctahedron, is shown by black solid lines. \textbf{b}~Projection to the trigonal plane. \textbf{c} The Fermi surface of holes centered at the T-point. This structure, dubbed ZONY~\cite{fauque2018}, consists of six interconnected pockets (shown in orange).}.
    \label{fig_SbFS}
\end{figure*}

\begin{table*}
\begin{center}
\begin{tabular}{|c||c|c|c|c|c|c|c|c|}
 \hline
Sample & Size (mm$^3$)& RRR  & $\mathrm{\rho_0}$ (n$\Omega$.cm)  & $\mathrm{\overline{s}}$ ($\mathrm{\mu}$m)& $\mathrm{\ell_0}$ ($\mathrm{\mu}$m) & $\rho_0$ $\mathrm{\overline{s}}$ (p$\Omega$ m$^{2}$) & A$_2$ (n$\Omega$.cm.K$^{-2}$) & B$_2$ (n$\Omega$.cm.K$^{-2}$)\\
\hline\hline
1 & ([0.25$\pm0.05 \times 0.5 \times 4.1$) & 260  & 159 & 350 & 17 & 0.56 & 0.70 $\pm$ 0.03 & 0.81 $\pm$ 0.05 \\
\hline
1b & ($0.2\times0.5 \times 4.6$) & 250 & 164 & 320 & 16 & 0.49 & 0.73 $\pm$ 0.04 & - \\
 \hline
2 & ($0.4 \times 0.4 \times 4.1$) &  430 & 94.6 & 400 & 28 & 0.38 & 0.56 $\pm$ 0.03 & 0.74 $\pm$ 0.03\\
\hline
3 & ($1.1\times 1.0 \times 10.0$) &  3000 & 13.4  & 1050 & 197 & 0.14 & 0.38 $\pm$ 0.03 & 0.68 $\pm$ 0.04 \\
\hline
3$^*$ & ($1.1\times 1.0 \times 7.0$) (cut from 3) & 3000  & 13.4 & 1050 & 197 & 0.14 & 0.38 $\pm$ 0.03 & - \\
 \hline
4 & ($1.0\times 5.0 \times 10.0$) &  1700 & 24.1  & 2240 & 110 & 0.54 & 0.32 $\pm$ 0.04 & 0.63 $\pm$ 0.08 \\
 \hline
5 & ($3.0\times 1.0 \times 10.0$) & 3700  & 11.1 & 1730 & 238 & 0.19 & 0.33 $\pm$ 0.03 & -\\
 \hline
6 & ($1.7\times1.8 \times$ 10.0) & 4200  & 9.8  & 1800 & 270 & 0.18 & 0.33 $\pm$ 0.03 & - \\
\hline
\end{tabular}
\caption{\textbf{Details of the samples.} Sb crystals used in this study were oriented along the bisectrix crystallographic axis. $\overline{s} = \sqrt{\mathrm{width} \times \mathrm{thickness}}$ represents the average diameter of the conducting cross-section. The residual resistivity ratio is defined as $RRR=\frac{\rho_{300K}}{\rho_0}$. The carrier mean free path $\ell_{0}$ was calculated from the residual resistivity and the expression for Drude conductivity assuming three spherical hole and three spherical electron pockets. This is a crude and conservative estimation, because the mean free path of hole-like and electron-like carriers residing in different valleys is likely to differ (See the Supplementary Note 2 for more details). Also given is the product of $\rho_0 \overline{s}$, a measure of crystalline perfection (Supplementary Note 2). The last two columns give the electrical ($A_2$) and thermal ($B_2$) $T^2$-resistivities prefactors.}
\label{table_size}
\end{center}
\end{table*}

\section{Results}

\subsection{The band structure}

Fig.\ref{fig_SbFS} shows the Fermi surface and the Brillouin Zone (BZ) of antimony~\cite{herrod1971,issi1979,Gonze1990,liu1995,fauque2018}. In this compensated semi-metal, electron pockets are quasi-ellipsoids located at the L-points of the BZ. The valence band crosses the Fermi level near the T-points of the Brillouin zone generating a multitude of hole pockets. The tight-binding picture conceived by Liu and Allen~\cite{liu1995}, which gives a satisfactory account of experimental data, implies that these pockets are not six independent ellipsoids scattered around the T-point~\cite{issi1979}, but a single entity~\cite{fauque2018} centered at the T-point formed by their interconnection (see Fig.\ref{fig_SbFS}.c).

One important point is that the pockets are small. The largest Fermi wave-vector is $0.22$ times the reciprocal lattice parameter~\cite{liu1995,fauque2018}. Since in an Umklapp collision between electrons, the sum of the Fermi wave-vectors should exceed the width of the BZ, Umklapp events cannot occur when $k_F<0.25$. The fact that the FS pockets are too small to allow Umklapp events will play an important role below.
\subsection{Electrical and thermal transport measurements}

\begin{figure}
\centering
\makebox{\includegraphics[width=0.45\textwidth]{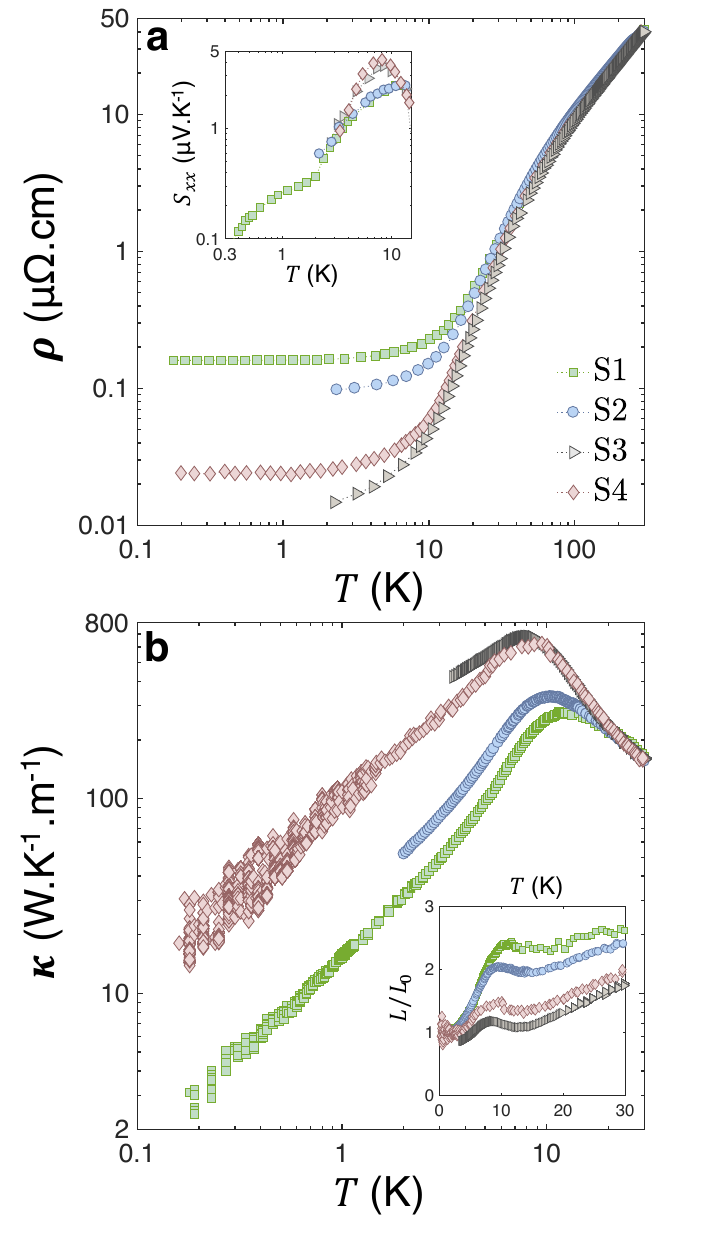}}
    \caption{\textbf{Zero-field transport properties. } \textbf{a} Electrical resistivity along the bisectrix direction, $\rho$, plotted as a function of temperature for the various sizes of Sb samples presented in Table \ref{table_size}. Inset shows the zero-field thermopower $S_{xx}$ as a function of the temperature of the same samples. \textbf{b} Temperature dependence of the thermal conductivity, $\kappa$, of the aforementioned Sb samples. Inset shows the Lorenz number $L$ plotted as $L/L_0$, where $L_0$ is the Sommerfeld number, as a function of temperature. $L/L_0=1$ corresponds to the recovery of the Wiedemann-Franz law.}
    \label{fig_Sb_fig1}
\end{figure}

All measurements were carried out using a conventional 4-electrode (two thermometers, one heater and a heat sink) setup (further details are given in the Method section). The Sb crystals are presented in Table \ref{table_size}. Electrical and heat currents were applied along the bisectrix direction of all samples. The electrical resistivity, shown in figure \ref{fig_Sb_fig1}.a, displays a strong size dependence below $T=25$K and saturates to larger values in the two thinner samples, as reported previously~\cite{Bogod1973}. As seen in Table \ref{table_size}, the mean free path remains below the average thickness, but tends to increase with the sample average thickness.

The thermal conductivity, $\kappa$, of the same samples is presented in figure \ref{fig_Sb_fig1}.b. $\kappa$ presents a peak whose magnitude and position correlates with sample size and resistivity. In large samples the peak is larger in amplitude and occurs at lower temperatures. Semi-metallic antimony has one electron and one hole for $\sim600$ atoms. The lattice and electronic contributions to the thermal conductivity are comparable in size. The inset of figure \ref{fig_Sb_fig1}.a shows the temperature dependence of the Seebeck coefficient in the same samples. The Seebeck coefficient remains below $5\mu$V/K, as reported previously~\cite{issi1979}, because of the cancellation between hole and electron contributions to the total Seebeck effect. The small size of the Seebeck response has two important consequences. First, it implies that the thermal conductivity measured in absence of charge current is virtually identical to the one measured in absence of electric field (which is the third Onsager coefficient~\cite{Behnia2015b}). The second is that the ambipolar contribution to the thermal transport is negligible and $\kappa= \kappa_{e}+ \kappa_{ph}$ (see the Supplementary Notes 5 and 6 respectively for a discussion of both issues).

The temperature dependence of the overall Lorenz number ($L=(\kappa\rho/T$)) divided by $L_0$, is plotted as a function of temperature in the inset of figure \ref{fig_Sb_fig1}.b. For $T<4$K, $L/L_0 \rightarrow 1$. The Wiedemann-Franz law is almost recovered below $T=4$K in all samples. At higher temperatures, $L$ displays a non-monotonic and size-dependent temperature dependence resulting from two different effects: a downward departure from the WF law in $\kappa_{e}$ and a larger share of $\kappa_{ph}$ in the overall $\kappa$.

\begin{figure}
\centering
\makebox{\includegraphics[width=0.45\textwidth]
    {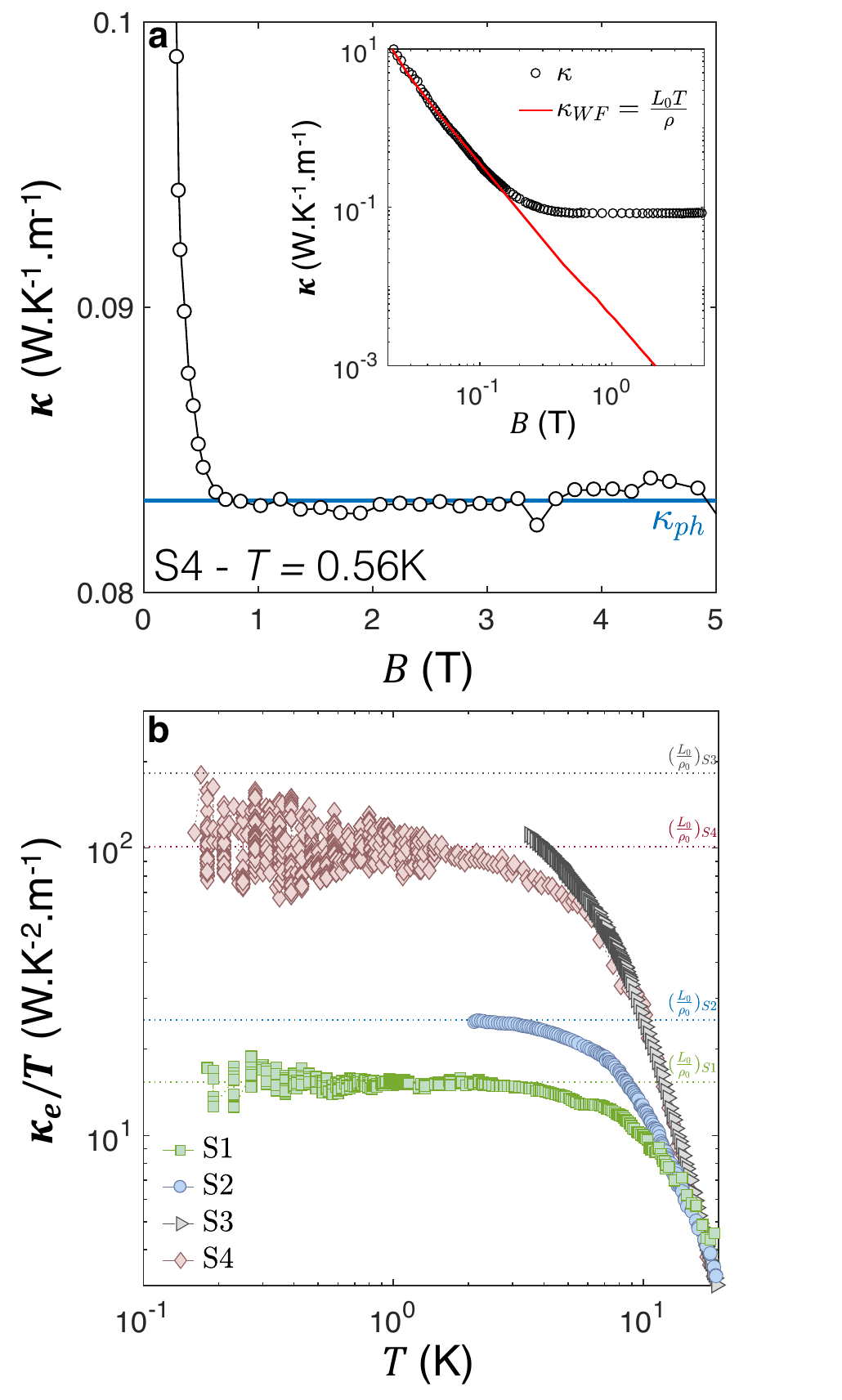}}
    \caption{\textbf{Using magnetic field to extract electronic and phononic components of thermal conductivity. } \textbf{a} Magnetic field dependence of the thermal conductivity of sample S4 at $T=0.56$K. The averaged field-independent fraction of $\kappa$, associated with the phonon contribution to $\kappa$ is shown as $\kappa_{ph}$. The inset shows a comparison of $\kappa$ and $\kappa_{WF}=\frac{TL_0}{\rho(B)}$ as a function of the magnetic field. For $B>0.5$T, the electronic thermal conductivity becomes negligible in regard of the phonon contribution. \textbf{b} Temperature dependence of the electronic thermal conductivity $\kappa_{e}=\kappa - \kappa_{ph}$ plotted as $\kappa_e/T$. Horizontal lines representing $L_0/\rho_0$ for the various samples are featured in the graph.}
    \label{fig_Sb_fig2}
\end{figure}

\begin{figure*}
\centering
\makebox{\includegraphics[width=1\textwidth]
    {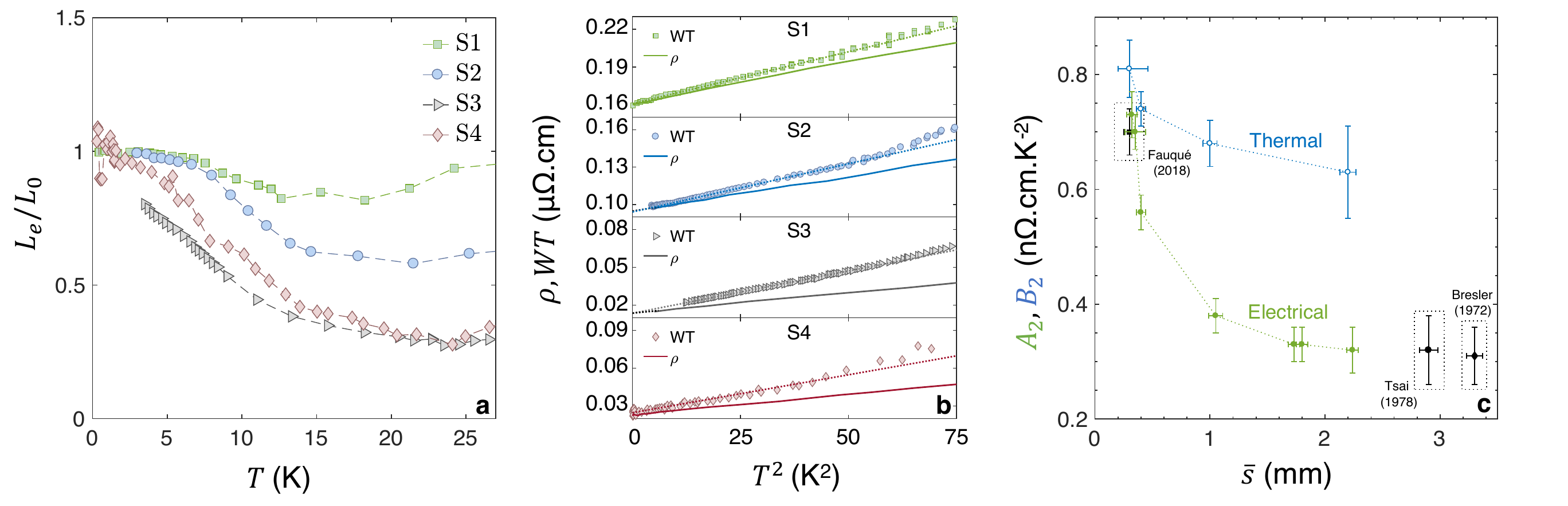}}
    \caption{ \textbf{The WF law, the $T$-square resistivities, and their evolution with disorder. } \textbf{a} Electronic fraction of the Lorenz number $L_e=\kappa_{e}\rho/T$ plotted as $L_{e}/L_0$, where $L_0$ is the Sommerfeld number, as a function of temperature. $L_e/L_0=1$ corresponds to the recovery of the Wiedemann-Franz law. \textbf{b} Thermal ($WT$) and electrical ($\rho$) resistivities plotted as functions of $T^2$ for the four sizes of Sb samples. $WT$ is featured as symbols while $\rho$ is shown as a solid line. All four graphs share a common x-axis and y-axis span. \textbf{c} Evolution of the electrical and thermal $T^2$-resistivities prefactors with sample size in Sb. Data points from \cite{fauque2018,tsai1978,bresler1972} are featured. Error-bars along the x-axis are defined by the uncertainty on the geometry of the samples while they are defined along the y-axis by the standard deviation deviation of the $T^2$-fit to the resistivity data.}
    \label{fig_Sb_fig3}
\end{figure*}

The application of a magnetic magnetic field provides a straightforward way to separate $\kappa_e$ and $\kappa_{ph}$ in a semi-metal with very mobile carriers~\cite{Uher1974}. Indeed, under the effect of a magnetic field, the electronic conductivity drastically collapses (the low-temperature magnetoresistance in Sb reaches up to $5.10^6\%$ at $B=1$T, as shown in Supplementary Figure 1) while the lattice contribution is left virtually unchanged. This is visible in the field dependence of $\kappa$, shown in figure \ref{fig_Sb_fig2}.a (for sample S4 at $T=0.56$K). One can see a sharp drop in $\kappa(B)$ below $B^*\approx0.5$T and a saturation at higher fields. The initial drop represents the evaporation of $\kappa_e$ due to the huge magnetoresistance of the system. The saturation represents the indifference of $\kappa_{ph}$ towards magnetic field. This interpretation is confirmed by the logarithmic plot in the inset and is further proven by the study of the low temperature thermal conductivity of Sb as a function of temperature under the effect of several fields presented in the Supplementary Figure 4. Below $B^*\approx0.1$T, $L_0T/\rho$ is close to $\kappa$, indicating that in this field window, heat is carried mostly by electrons and the WF law is satisfied. However, by $B^*\approx 1$T, $L_0T/\rho$ is three orders of magnitude lower than $\kappa$, implying that at this field, heat is now mostly carried by phonons with a vanishing contribution from electrons. The electronic component of thermal conductivity separated from the total thermal conductivity, ($\kappa_e(T)=\kappa(B=0)(T)-\kappa(B=1T)(T)$) is shown in figure \ref{fig_Sb_fig2}.b. One can see that, for all four samples and at sufficiently low temperature, $\kappa_e/T$ becomes constant (and equal to $L_0/\rho_0$). It is the subsequent downward deviation at higher temperatures which will become the focus of our attention. We construct the electronic Lorenz ratio $L_e=\kappa_e\rho/T$ and show its evolution with temperature in figure \ref{fig_Sb_fig3}.a. Below $T<4$K, $L_e\simeq L_0$ in all samples, save for S3, the cleanest. With increasing temperature, $L_e/L_0$ dives down and the deviation becomes larger as the samples become cleaner.

\begin{figure}
\makebox{\includegraphics[width=0.35\textwidth]
    {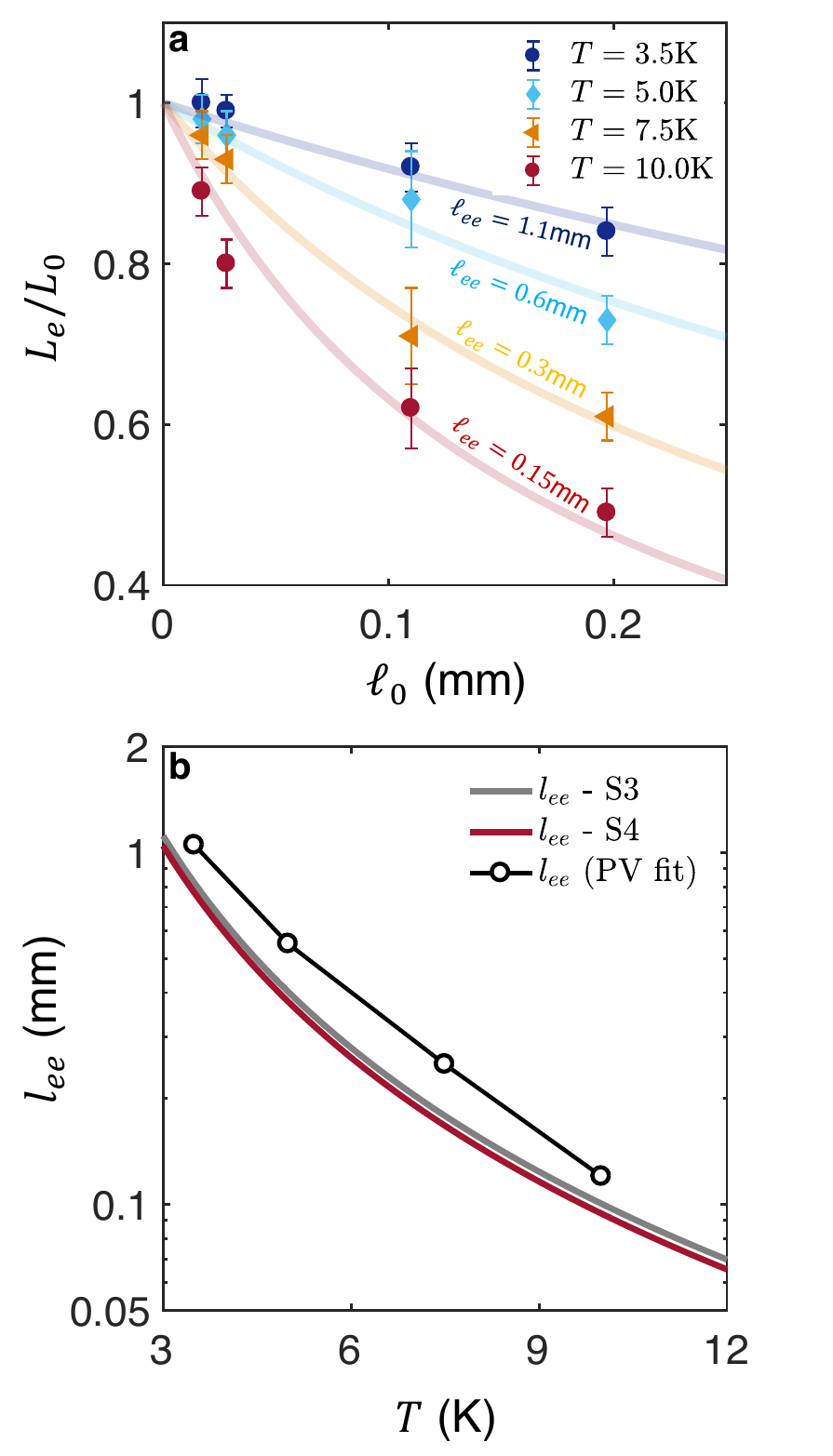}}
    \caption{\textbf{The evolution of the Wiedemann-Franz correlation with the ratio of momentum-relaxing and momentum conserving  mean free paths.} \textbf{a}~The electronic Lorenz number $L_e$ at $T=10$K, normalized by the Sommerfeld value $L_0$, plotted as a function of the residual mean free path $\ell_0$ at various temperatures. The solid lines correspond to a fit given by the equation $L/L_0=\frac{1}{1+\frac{\ell_0}{\ell_{ee}}}$ proposed by Principi and Vignale (PV) ~\cite{principi2015}. $\ell_0$ refers to the zero-temperature Drude mean free path while $\ell_{ee}(T)$ is the typical distance traveled by a charge carrier in-between two momentum-conserving collisions. Error bars are defined from the experimental uncertainty on $L_e$ featured in Fig.\ref{fig_Sb_fig3}.a. \textbf{b}~Comparison of $l_{ee}$ determined by the fit to the aforementioned PV formula and what is yielded by assuming that the difference between the two $T$-square resistivities represents the fraction of collisions which conserve momentum. In that case,
    $\ell_{ee}=\frac{\ell_0\rho_0}{(B_2-A_2)T^2}$.}
    \label{fig_Sb_fig6}
\end{figure}

Let us scrutinize separately the temperature dependence of the electrical and the thermal resistivities. The latter can be expressed in the familiar units of resistivity (i.e. $\Omega$.m), using  $WT = L_0T/\kappa_{e}$ as a shorthand. Figure \ref{fig_Sb_fig3}.b shows $\rho$ and $WT$  as a function of $T^2$ for the four different samples. In the low-temperature limit, an asymptotic $T^2$ behavior is visible in all samples and the two lines corresponding to $\rho$ and $WT$ have identical y-axis intercepts, thus confirming the recovery of the WF Law in the zero-temperature limit. In every case, the slope of $WT(T^2)$ is larger than that of $\rho(T^2)$, indicating that the prefactor of the \textit{thermal} $T$-square resistivity (dubbed $B_2$) is larger than the prefactor of the \textit{electrical} $T$-square resistivity (dubbed $A_2$). This behavior, observed for the first time in Sb, was previously reported in a handful of metals, namely  W~\cite{wagner1971}, WP$_2$~\cite{jaoui2018}, UPt$_3$~\cite{lussier1994} and CeRhIn$_5$~\cite{paglione2005}.

\section{Discussion}
$T$-square resistivity arises due to $e-e$ collisions. In the momentum-relaxing picture, the common explanation for the experimentally observed $B_2>A_2$ inequality is the under-representation of small-angle scattering in the electrical channel, which damps the electric prefactor $A_2$, but not its thermal counterpart $B_2$~\cite{ziman1972,wagner1971,paglione2005,jaoui2018,li2018}. This picture cannot explain that, as seen in Figure \ref{fig_Sb_fig3}.b, the two slopes are further apart in the cleaner samples. The evolution of the two prefactors with sample dimensions is presented in figure \ref{fig_Sb_fig3}.c. The figure also includes previous data on the slope of electrical $T^2$-resistivity~\cite{fauque2018,tsai1978,bresler1972}. One can see the emergence of a consistent picture: the electrical ($A_2$) prefactor displays a significant size dependence and the $A_2/B_2$ ratio substantially decreases with the increase in sample size and electronic mean free path. 

Because of momentum conservation, $e-e$ collisions cannot decay the momentum flow by themselves. Such collisions can relax momentum through two mechanisms known as Umklapp and interband (or Baber) scattering. There are two known cases of $T$-square resistivity in absence of either mechanisms~\cite{lin2015,Wang2020}.

The smallness of the Fermi surface in Sb excludes the Umklapp mechanism. However, the interband mechanism is not excluded. It can generate both a $T$-square and a $A_2/B_2$ ratio lower than unity~\cite{li2018}.~Li and Maslov~\cite{li2018} have argued that the ratio of the two prefactors (and therefore the deviation from the WF law) in a compensated semi-metal like Sb is tuned by two material-dependent parameters: i) the screening length and ii) the relative weight of interband and intraband scattering. In their picture, increasing the screening length would enhance $B_2$ and leave $A_2$ unchanged. Enhancing interband scattering would also reduce the Lorenz ratio. Given that neither of these two is expected to change with the crystal size or imperfection, the evolution seen in figure \ref{fig_Sb_fig3}.c cannot be explained along either of these two lines.

In contrast, the hydrodynamic picture provides a straightforward account of our observation. The Principi and Vignale scenario~\cite{principi2015} predicts that the deviation from the WF law should become more pronounced with increasing carrier lifetime (or equivalently mean free path $\ell_0$): $L_e/L_0=1/(1+\ell_0/\ell_{ee})$. Such a picture provides a reasonable account of our observation, as seen in figure \ref{fig_Sb_fig6}.a, which shows the variation of $L_e/L_0$ at different temperatures with carrier mean free path. In this picture, the evolution of the Lorenz ratio with $\ell_0$ would imply a mean free path for MC $e-e$ scattering, $\ell_{ee}$, which ranges from $0.15$mm at $T=10$K to $1.1$mm at $T=3.5$K. 

These numbers are to be compared with $\ell_{ee}$ extracted from the magnitude of ($B_2$, $A_2$), assuming that MC $e-e$ collisions generate the difference between these two quantities and the Drude formula. As seen in figure \ref{fig_Sb_fig6}.b, while the two numbers closely track each other between $T=3$K and $T=10$K, a difference is found. $\ell_{ee}$ extracted from the isotropic Drude formula is $1.6$ times smaller than the one yielded by the isotropic Principi-Vignale formula. Now, the electronic structure of antimony is strongly anisotropic with a tenfold difference between the longest and the shortest Fermi wave-vectors along different orientations~\cite{liu1995}. In such a context, one expects an anisotropic $\ell_{ee}$, with different values along different orientations. Moreover, intervalley scattering between carriers remaining each in their only valley and scattering between electrons and holes should also have characteristic length scales. Therefore, the present discrepancy is not surprising and indicates that at this stage, only the order of magnitude of the experimental observation is accounted for by a theory conceived for isotropic systems~\cite{principi2015}. Note the macroscopic ($\sim$mm) magnitude of $\ell_{ee}$ near $T\sim4$K which reflects the fact that electrons are in the ultra-degenerate regime ($T/T_F \sim 4 \times 10^{-3}$) and therefore, the distance they travel to exchange momentum with another electron is almost six orders of magnitude longer than the distance between two electrons.

An account of boundary scattering is also missing. The decrease in $\rho_0$ with sample size in elemental metals have been widely documented and analyzed by pondering the relative weight of specular and diffusive scattering~\cite{sambles1980}. This can also weigh on the magnitude of $A_2$~\cite{maas1985}. However, a quantitative account of the experimental data, by employing Soffer's theory~\cite{soffer1967}, remains unsuccessful ~\cite{Bogod1973,maas1985,sambles1983}. The role of surface roughness acquires original features in the hydrodynamic regime~\cite{kiselev2019}, which are yet to be explored by experiments on samples with mirror/matt surface dichotomy.

\begin{table*}[htbp]
\begin{center}
\begin{tabular}{|c|c||c|c|c|c|c|c|}
 \hline
System & Density (cm$^{-3}$) & T$_F$ (K)  & k$_F$ (nm$^{-1}$) & $\kappa T|_0$ (W.m$^{-1}$)& $\frac{\mathrm{E_F^2 k_F}}{\hbar}$ (W.m$^{-1}$) & B$_0$ & $\tau_{\kappa}$T$^2$(s.K$^2$)\\
\hline\hline
$^3$He~\cite{greywall1984} &$1.63\times 10^{22}$& 1.8 & 7.8 & 2.9 $\times 10^{-4}$& 0.04 & 137  &  3.9 $\times 10^{-13}$ \\
\hline
Sb &$n+p=1.1\times 10^{20}$& 1100~\cite{liu1995} &  0.8 (average)~\cite{liu1995} & 3900-7900   & 1500 & 0.4-0.8 & 1.5-3.0 $\times 10^{-8}$   \\
\hline
\end{tabular}
\caption{\textbf{ Comparison of two Fermi liquids.} The density of atoms at ambient pressure in $^3$He is two orders of magnitude larger than the total density of electron-like and hole-like carriers $(n+p)$ in Sb. Also listed in this Table are the average Fermi temperature, the average Fermi momentum and the magnitude of the experimentally-resolved $\kappa T|_0$ (W.m$^{-1}$). Its natural units are $\frac{E_F^2 k_F}{\hbar}$. $B_0$ is defined in the main text. $\tau_kT^2$ quantifies the rate of fermion-fermion collisions.}
\label{table_FLs}
\end{center}
\end{table*}

Having shown that the experimentally-resolved $T^2$ resistivity is (at least) partially caused by thermal amplification of momentum exchange between fermionic quasiparticles, we are in a position to quantify $\kappa T|_0$ in antimony and compare it with the case of $^3$He.

Its lower boundary is $L_0/(A_2-B_2)$ and the upper boundary $L_0/B_2$. This yields $3900 < \kappa T|_0 < 7900$ in units of W.m$^{-1}$. This is six orders of magnitude larger than in normal liquid $^3$He~\cite{greywall1984} (see Table \ref{table_FLs}). Such a difference is not surprising since: i) $\kappa T|_0$ of a Fermi liquid is expected to scale with the cube of the Fermi momentum ($p_F$) and the square of the Fermi velocity ($v_F$)~\cite{calkoen1986}; and ii) $^3$He is a strongly correlated Fermi liquid while Sb is not. More specifically $\kappa T|_0$ can be written in terms of the Fermi wave-vector ($k_F$) and the Fermi energy ($E_F$):

\begin{equation}
\kappa T|_0= \frac{1}{B_0}\frac {E_F^2k_F} {\hbar}
\label{kappaT}
\end{equation}

This equation is identical to equation 17 in ref~\cite{calkoen1986}. The dimensionless parameter $B_0$ (See Supplementary Note 7 for more details) quantifies the cross section of fermion-fermion collisions.

In the case of $^3$He, measuring the temperature dependence of viscosity~\cite{bertinat1974,alvesalo1975} leads to $\eta T^2_0$ and measuring the temperature dependence of thermal conductivity~\cite{greywall1984} leads to $\kappa T|_0$. The rate of fermion-fermion collisions obtained  with these two distinct experimental techniques are almost identical : $\tau_{\eta}T^2\approx\tau_{\kappa}T^2$~\cite{alvesalo1975}. Calkoen and  van Weert~\cite{calkoen1986} have shown that the agreement between the magnitude of $\kappa T|_0$, the Landau parameters and the specific heat~\cite{greywall1983} is of the order of percent. 

$^3$He is a dense strongly-interacting quantum fluid, which can be solidified upon a one-third enhancement in density. As a consequence, $B_0 \gg 1$. In contrast, the electronic fluid in antimony is a dilute gas of weakly interacting fermions and $B_0$ is two orders of magnitude lower, as one can see in Table~\ref{table_FLs}. The large difference in $B_0$ reflects the difference in collision cross section caused by the difference in density of the two fluids. 

The $T^2$ fermion-fermion scattering rate can be extracted and $\tau_{\kappa}T^2$ can be compared with the case of $^3$He~\cite{alvesalo1975,bertinat1974,greywall1984,wolfle1979} (See Table~\ref{table_FLs}). As expected, it is many orders of magnitude smaller in Sb than in its much denser counterpart. A similar quantification is yet to be done in strongly-correlated electronic fluids.

In summary, we found that the ratio of the thermal-to-electrical $T$-square resistivity evolves steadily with the elastic mean free path of carriers in bulk antimony. The momentum-conserving transport picture provides a compelling explanation for this observation. In this approach, thermal resistivity is in the driver's seat and generates a finite electrical resistivity which grows in size as the sample becomes dirtier.

This a hydrodynamic feature, since the same fermion-fermion collisions, which set momentum diffusivity (that is viscosity) set energy diffusivity (the ratio of thermal conductivity to specific heat). Note that this is a feature specific to quantum liquids, in contrast to the upward departure from the WF law reported in graphene when carriers are non-degenerate~\cite{crossno2016}.

The observation of this feature in Sb was made possible for a combination of properties. i) The mean free path of carriers was long enough to approach the sample dimensions; ii) The Normal collisions outweigh Umklapp collisions because the Fermi surface radii of all pockets is less than one-fourth of the width of the Brillouin zone. Finally, at the temperature of investigation scattering by phonons is marginal. All these conditions can be satisfied in low-density semimetals such as Bi~\cite{collaudin2015} or WP$_2$~\cite{gooth2018,jaoui2018}. In contrast, in a high-density metal such as PdCoO$_2$~\cite{moll2016}, such a feature is hard to detect. Note only, due to the large Fermi energy, the $T$-square resistivity is small and undetectable~\cite{Hicks2012}, but also due to the large Fermi radius~\cite{Mackenzie_2017}, electron-electron collisions are expected to be mostly of Umklapp type.

Beyond weakly correlated semi-metals, our results point to a novel research horizon in the field of strongly correlated electrons. One needs quasi-ballistic single crystals (which can be provided thanks to Focused-Ion-Beam technique) of low-density correlated metals. URu$_2$Si$_2$~\cite{Kasahara2007} and PrFe$_4$P$_{12}$~\cite{Pourret2006}, known to be low-density strongly correlated Fermi liquids, appear as immediate candidates but other systems may qualify. The electron-electron collision cross section, which can be quantified by a study similar to ours should be much larger than what is found here for a weakly correlated system such as Sb.

\section{Methods}
\subsection{Samples}
Sb crystals were commercially obtained through MaTeck GmbH. Their dimensions are given in Table 1 of the main text. Samples S1, S1b and S2 were cut from a ingot of Sb using a wire saw. Samples S3, S4, S5 and S6 were prepared by MaTeck to the aforementioned dimensions: sample S4 was cut while samples S3, S5 and S6 were etched to these dimensions. Sample S3 was measured before and after a cut of a few mm perpendicular to the bisectrix direction. The long axis of all samples were oriented along the bisectrix direction. 
\subsection{Measurements}
The thermal conductivity measurements were performed with a home-built one-heater-two-thermometers set-up. Various thermometers (Cernox chips 1010 and 1030 as well as RuO$_2$) were used in this study. Our setup was designed to allow the measurement of both the thermal conductivity, $\kappa$ and the electrical resistivity, $\rho$ with the same electrodes.\\ The thermometers were either directly glued to the samples with Dupont 4922N silver paste or contacts were made using $25\mu$m-diameter silver wires connected to the samples via silver paste (Dupont 4922N). Contact resistance was inferior to $1\Omega$. The thermometers were thermally isolated from the sample holder by manganin wires with a thermal conductance several orders of magnitude lower than that of the Sb samples and silver wires.
The samples were connected to a heat sink (made of copper) with Dupont 4922N silver paste on one side and to a RuO$_2$ chip resistor serving as a heater on the other side. Both heat and electrical currents were applied along the bisectrix direction. The heat current resulted of an applied electrical current $I$ from a DC current source (Keithley 6220) to the RuO$_2$ heater. The heating power was determined by $I\times V$ where $V$ is the electric voltage measured across the heater by a digital multimeter (Keithley 2000). The thermal conductivity was checked to be independent of the applied thermal gradient by changing $\Delta T/T$ in the range of 10\%. Special attention was given not to exceed $\Delta T/T\vert_{max}=10\%$. \\The thermometers were calibrated \textit{in-situ} during each experiment and showed no evolution with thermal cycling. Special attention was given to suppress any remanent field applied to the sample and self-heating effects. \\The accuracy of our home-built setup was checked by the recovery of the Wiedemann-Franz law in an Ag wire at $B=0$T and $B=10$T through measurements of the thermal conductivity and electrical resistivity. At both magnetic fields, the WF was recovered at low temperatures with an accuracy of $1\%$ ~\cite{jaoui2018}.

\section{Competing interests}
The authors declare no conflict of interest.

\section{Author contributions}
A.J carried out the experiments. A.J, B.F and K.B analyzed the data and wrote the manuscript.

\section{Acknowledgement}
This work is supported by the Agence Nationale de la Recherche (ANR-18-CE92-0020-01; ANR-19-CE30-0014-04) and by Jeunes Equipes de l$'$Institut de Physique du Coll\`ege de France.

\section{Data availability}
All data supporting the findings of this study are available from the corresponding authors upon request.
\clearpage
\onecolumngrid
\section*{Supplementary Material for 'Thermal resistivity and hydrodynamics of the degenerate electron fluid in antimony'}

\setcounter{equation}{0}
\setcounter{figure}{0}
\setcounter{table}{0}
\setcounter{section}{0}
\setcounter{page}{1}
\makeatletter

\section{Supplementary Figures}

\begin{suppfigure}[H]
\centering
\makebox{\includegraphics[width=1.0\textwidth]
    {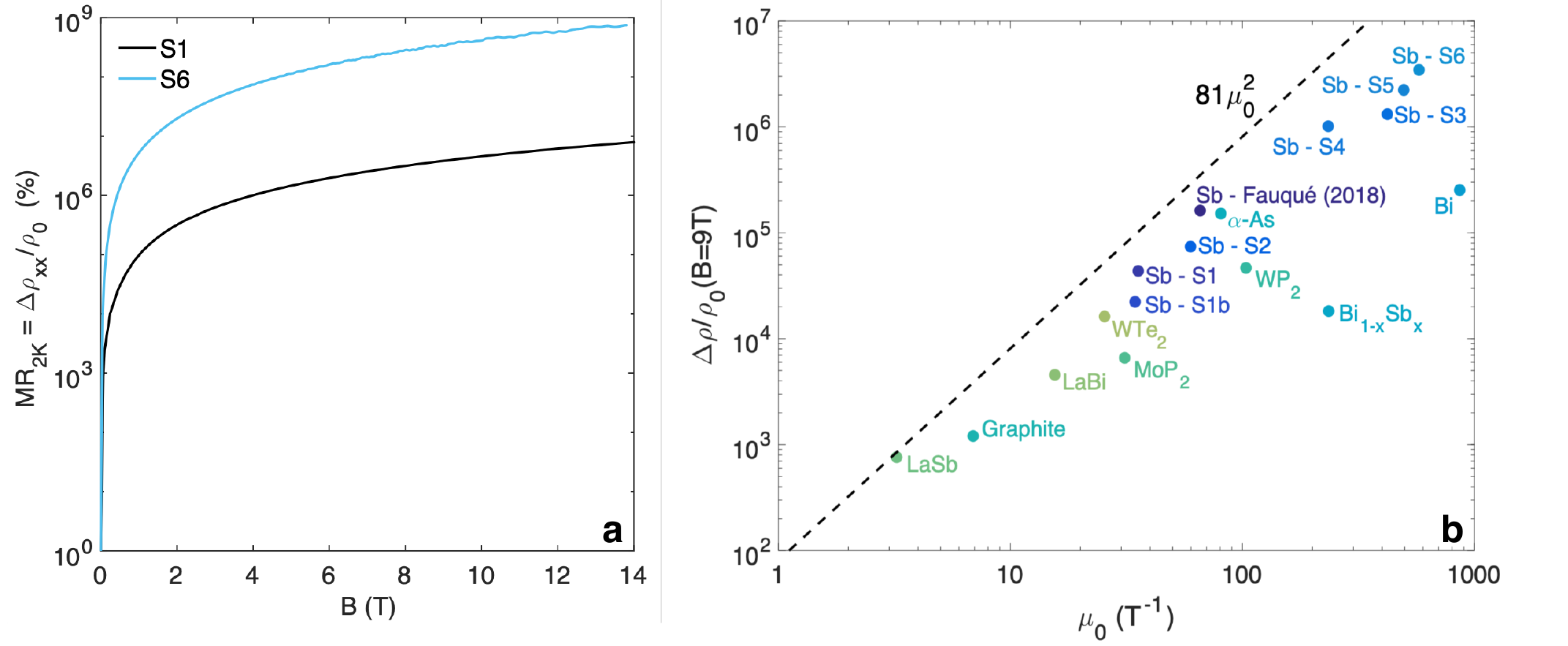}}
    \caption{: \textbf{Magnetoresistance of Sb. a}~Magnetoresistances of sample S1 and S6 at $T=2$K. \textbf{b}~Magnetoresistance of various semi-metals at $B=9$T and $T=2$K as a function of the mobility $\mu_0=1/(\rho_0(n+p)e)$ where $e$ is the elementary charge, $n$ and $p$ are the electron and hole densities and $\rho_0$ the zero field resistivity at $T=2$K. $\mu_0$ is expressed in Tesla$^{-1} =10^4 $cm$^2$.V$^{-1}$.s$^{-1}$}
    \label{fig_sup_MR}
\end{suppfigure}

\begin{suppfigure}[H]
\centering
\makebox{\includegraphics[width=0.4\textwidth]
    {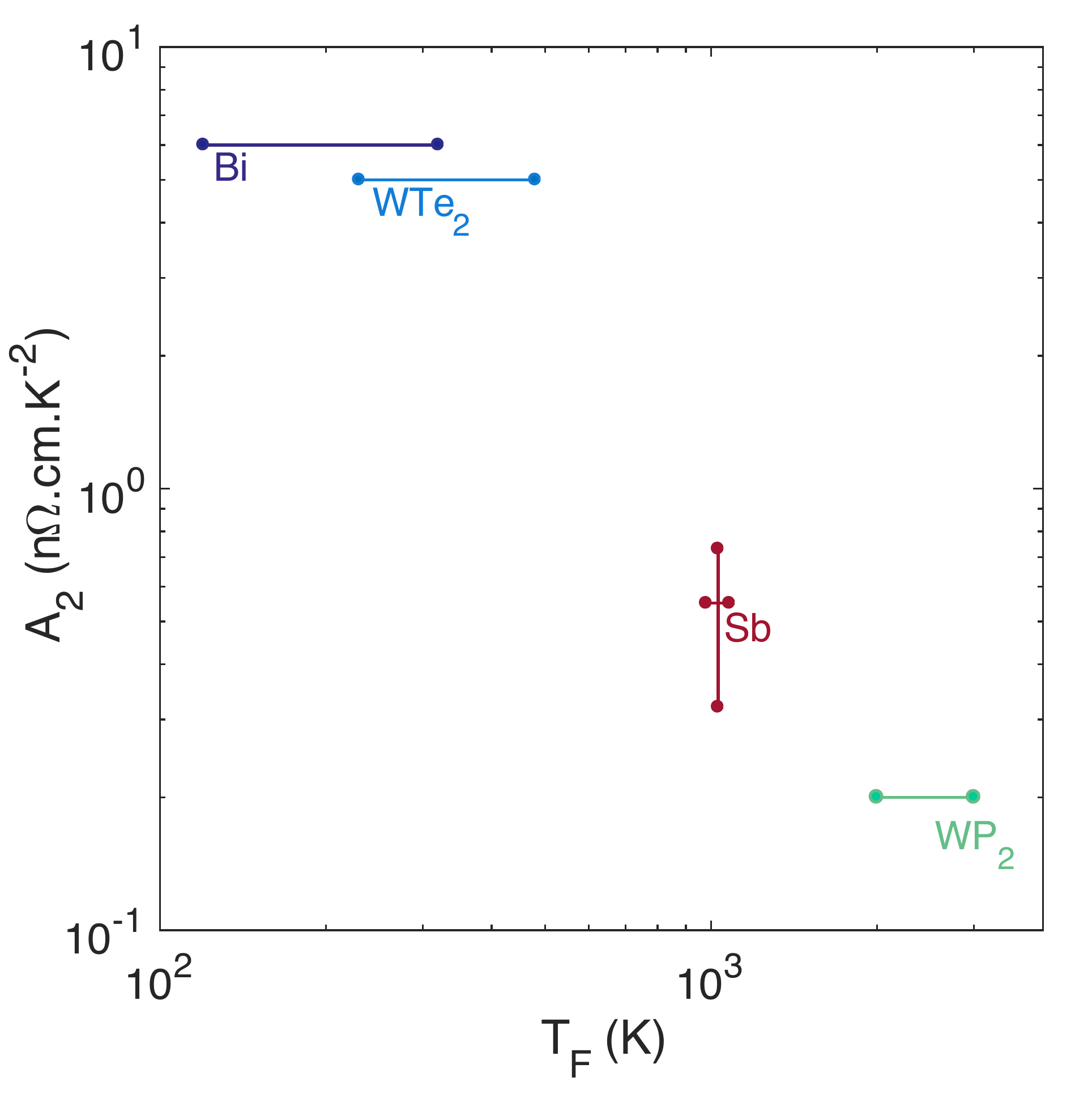}}
    \caption{: \textbf{Amplitude of the $T^2$-resistivity prefactor in semi-metals. a}~ Electrical $T^2$-resistivity prefactor ($A_2$) plotted as a function of the Fermi temperature for the semi-metals discussed in table \ref{tab_semimetals}. $T_F$ was taken for both electrons and holes for Sb~\cite{issi1979,liu1995}, Bi~\cite{edelman1976,liu1995}, WTe$_2$~\cite{zhu2015} and WP$_2$~\cite{schonemann2017,kumar2017} and the prefactor $A_2$ for Sb was taken from this work, Bi \cite{issi1979}, WTe$_2$ \cite{zhu2015} and WP$_2$ \cite{jaoui2018}.}
    \label{Sb_supp_TF_A2}
\end{suppfigure}

\begin{suppfigure}[H]
\centering
\makebox{\includegraphics[width=\textwidth] {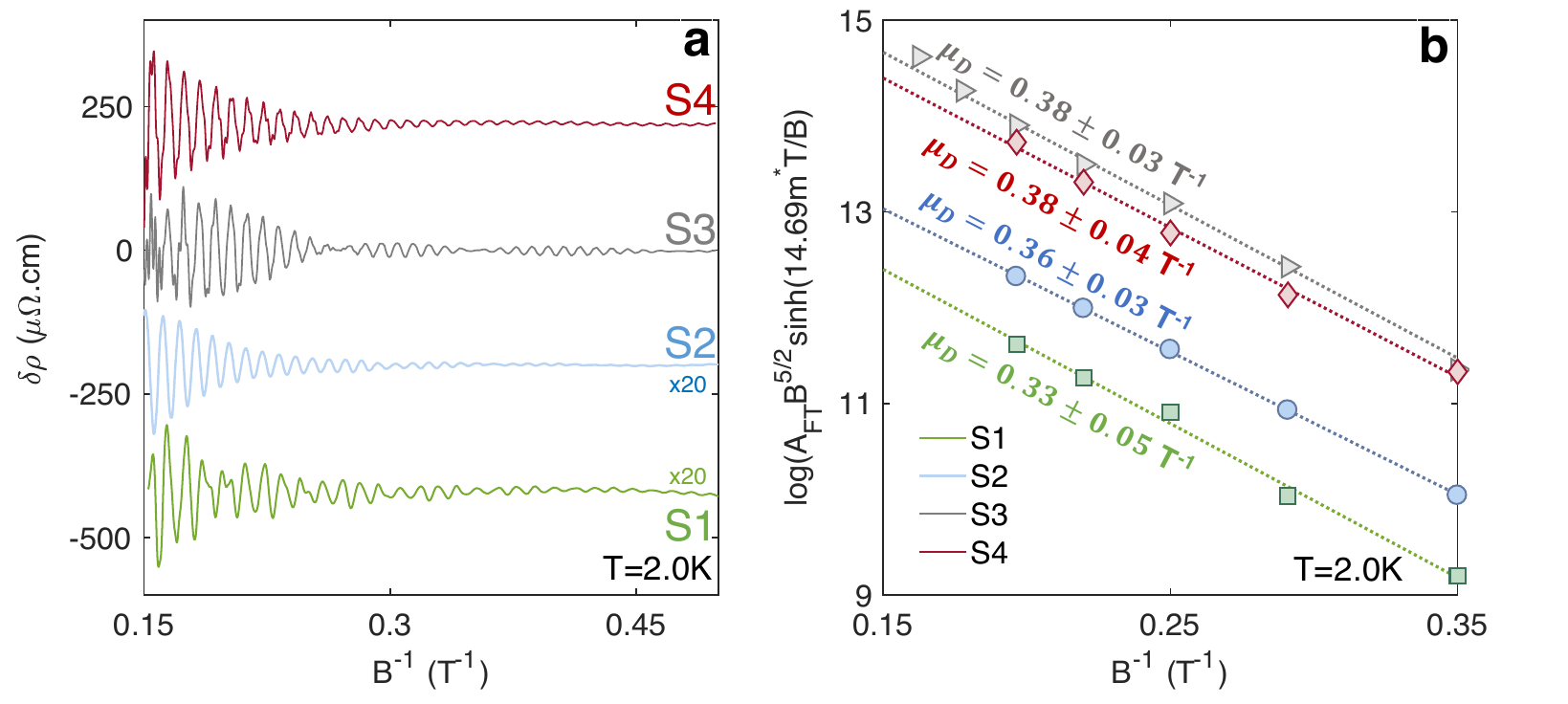}}
    \caption{: \textbf{Effect of sample size on the quantum oscillations observed in Sb. a}~Quantum oscillations of the magnetoresistance (the Shubnikov-de Haas effect) in four Sb samples as listed in table 1 of the main text. In all cases, the field was applied along the trigonal axis and the current was applied along the bisectrix axis. \textbf{b}~Dingle analysis of the data revealing a quasi-identical mobility.}
    \label{fig_Dingle}
\end{suppfigure}

\begin{suppfigure}[H]
\centering
\makebox{\includegraphics[width=0.5\textwidth]
    {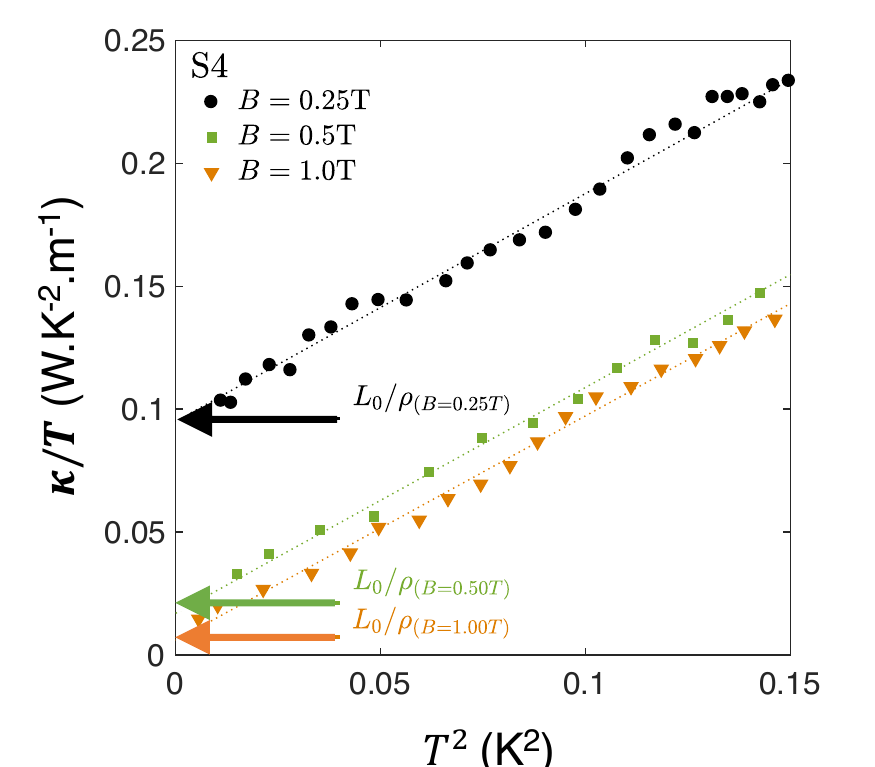}}
    \caption{: \textbf{Recovery of Wiedemann-Franz law at small applied magnetic fields. a}~Thermal conductivity plotted as $\kappa/T$ as a function of $T^2$ in sample S4 for three successive magnetic fields. The symbols show the experimental values of $\kappa/T$ and the dotted lines are the linear fit while the value of $L_0/\rho_{B}$ are featured as arrows. The recovery of the Wiedemann-Franz law at each magnetic field is shown by the intercept of the dotted lines and arrows.}
    \label{fig_sup_WF_B}
\end{suppfigure}

\section{Supplementary Tables}

\begin{supptable}[H]
\makebox[\columnwidth][c]{\begin{tabular}{|c||c|c|c|c|c|c|c|}
\hline
 Semi-metal & $n=p$ (cm$^{-3}$) & $\overline{\rho_0}$ ($\mu\Omega$.cm) & $\overline{\mu_0}$ (m$^2$.V$^{-1}$.s$^{-1}$) & $m^*$ ($m_0$) & $\overline{T_{F,e}}$ (K)& $\overline{T_{F,h}}$ (K) & References \\
 \hline \hline
    Sb & $5.5\times 10^{19}$ & $\sim 0.05$ & $\sim 500$ & $0.07-1$ & $1080$ & $980$ & \cite{issi1979,liu1995} \\\hline
    Bi & $3.0\times 10^{17}$ & $\sim 1$ & $\sim 1000$ & $0.001-0.612$ & $320$ & $120$ & \cite{edelman1976,liu1995}  \\\hline
    WP$_2$ & $2.5\times 10^{21}$ & $\sim0.005$ & $\sim 400$ & $0.7-1.9$ & $3000$ & $2000$ & \cite{schonemann2017,kumar2017}  \\\hline
    WTe$_2$ & $6.8\times 10^{19}$ & $\sim 1$ & $\sim 5$ & $0.1-1.2$ & $480$ & $230$ & \cite{zhu2015} \\\hline
   \end{tabular}
   }
\caption{: \textbf{Comparison of prominent semi-metals.} We compare the electronic concentration $n$, typical residual resistivity $\overline{\rho_0}$, typical electronic mobility $\overline{\mu_0}$, carriers effective mass $m^*$ and typical Fermi temperature of electrons $\overline{T_{F,e}}$ and holes $\overline{T_{F,h}}$ in these materials. References used to construct this table are featured in the last column.}
\label{tab_semimetals}
\end{supptable}

\begin{supptable}[H]
\begin{center}
\begin{tabular}{|c|c|c|c|c|c|}
 \hline
Sample & $\rho_0$(n$\Omega$ cm) & $\mu_0$ (m$^2$V$^{-1}$s$^{-1}$)&  $\mu_D$ (m$^2$V$^{-1}$s$^{-1}$)& r\\
\hline\hline
1 & 159 & 71 & 0.33 & 215 \\
\hline
2 &  94.6 & 120 & 0.36 & 333 \\
\hline
3 & 13.4  & 848 & 0.38 & 2231\\
\hline
4 &  24.1 & 772 & 0.38 & 2031 \\
\hline
\end{tabular}
\caption{: \textbf{The two mobilities in four different crystals}. Transport mobility, $\mu_0$ has been extracted from $\rho_0$ using $\mu_0=1/\rho_0e(n+p)$ and Dingle mobility $\mu_D$ is extracted from a Dingle analysis of the quantum oscillations. $r$ is the ratio of the two mobilities.}
\label{table_size_sup}
\end{center}
\end{supptable}
\section{Supplementary Notes}
\subsection{Supplementary Note 1 : Magnetoresistance and mobility.}
The high mobility of charge carriers in Sb leads to a very large magnetoresistance, as reported in Ref.\cite{fauque2018}. The samples presented in this study confirm this. As an example, the magnetoresistance of sample S6 at $T=2$K and $B=9$T is shown in Supplementary Figure \ref{fig_sup_MR}.a. This large magnetoresistance translates into a suppression of the electronic thermal conductivity through the Wiedemann-Franz law. As a consequence, the separation of lattice and electronic contributions of $\kappa$ becomes straightforward. The mobility and the magnetoresistance of the samples used in this study are shown in Supplementary Figure \ref{fig_sup_MR}.b and compared to other semi-metals. One can see that carriers in Sb are extremely mobile compared to most other semi-metals.\\ Supplementary Table \ref{tab_semimetals} compares the electronic properties of Sb with a few other semi-metals. Supplementary Figure \ref{Sb_supp_TF_A2} shows the magnitude of the electrical $T^2$-resistivity prefactor $A_2$ in four different semi-metals. One can see that $A_2$ decreases with increasing Fermi temperature, as previously noted in the case of numerous dilute metals~\cite{collignon2019}. The correlation between $A_2$ and $E_F^2$ is an extension of the Kadowaki-Woods correlation~\cite{tsujii2003} to low-density systems~\cite{Wang2020}.

\subsection{Supplementary Note 2 : Estimation of the electronic mean-free-path}

In the Drude picture, the measured residual resistivity, $\rho_0$ is related to the scattering time of electrons and holes and their masses by Supplementary Eq.(\ref{eq_Drude}): 
\begin{equation}
    \rho_0^{-1}=e^2 (\frac{n\tau_e}{m^*_e}+\frac{p\tau_h}{m^*_h})
    \label{eq_Drude}
\end{equation}
In Sb, the compensation between electron and hole densities holds with an accuracy of $10^{-4}$ and one has: $n=p=5.5\times 10^{19}$ cm$^{-3}$~ \cite{fauque2018}. However, electrons and hole pockets have different shapes, significant mass anisotropy and are not aligned parallel to each other. Their associated scattering time is unlikely to be identical.

The mean-free-path of the samples given in table 1 of the main text was extracted from their residual resistivity using a conservative and crude approximation. If the Fermi surface is composed of $z_h$ spheres for hole-like and $z_e$ spheres for electron-like carriers, then the average Fermi wave-vector for both $k_F^e=(3\pi^2(n/z_e))^{1/3}$ and $k_F^h=(3\pi^2(p/z_h))^{1/3}$.

Now neglecting the possibility that for holes the valleys may be connected to each other (Fig.2 of the main text), we took $z_e=z_h=3$ and found $k_F^h= k_F^e=0.82$nm$^{-1}$. Depending on the orientation, the actual and anisotropic $k_F$ resides between $0.45$ and $2.4$nm$^{-1}$~\cite{liu1995}. In this approximation, the mean-free-path can be evaluated using the Drude formula and becomes Supplementary Eq.(\ref{eq_mfp}):

\begin{equation}
    \ell_0=\frac{3\pi}{2(z_e+z_h)}\frac{1}{\rho_0}\frac{h}{e^2}\frac{1}{k_F^2}
    \label{eq_mfp}
\end{equation}

The $\ell_0$ values given given in Table 1 of the main text, has been extracted using this equation with $z_e=z_h=3$. In this approximation, the residual resistivity times the average diameter $\overline{s}$ has a lower boundary set by the carrier concentration (Supplementary Eq.(\ref{eq_finite size})).  
\begin{equation}
   (\rho_0\overline{s})_{min} =\frac{3\pi}{2(z_e+z_h)}\frac{h}{e^2}\frac{1}{k_F^2}) 
    \label{eq_finite size}
\end{equation}

Putting $z_e=z_h=3$, and $n=p=5.5\times 10^{19}$ cm$^{-3}$, one finds $(\rho_0\overline{s})_{min}= 0.03$p$\Omega$.m$^2$. The lowest reported value reported in the scientific literature for a Sb crystal is ($\rho_0\overline{s}\approx 0.1$p $\Omega$.m$^2$~\cite{hatzopoulos1985}), slightly lower than our best Sb crystal (S3) ($\rho_0\overline{s}=0.14 $p $\Omega$.m$^2$).

\subsection{Supplementary Note 3 : Dingle mobility}

Quantum oscillations have been used to map the Fermi surface of Sb~\cite{herrod1971}. As seen in Supplementary Figure \ref{fig_Dingle}.a, they are easily observable in our crystals. The Dingle analysis yields a mobility, which is much lower than the mobility extracted from residual resistivity. Moreover, as one can see in in Supplementary Figure \ref{fig_Dingle}.b, they barely change in four different samples, in spite of their ten-fold variation in residual resistivity. While $\ell_0$ in sample S1 is 10 times shorter than in sample S3, the mobility is only $1.2$ time larger.

Such a large discrepancy have been found in other dilute metals~\cite{Liang2015,kumar2017}. In all three cases, the quasi-particle lifetime  extracted from  transport is orders of magnitude longer than the Dingle scattering time. Our cleanest samples show a $2000$-fold discrepancy, which is to be compared to what was reported for the cleanest sample in Cd$_2$As$_3$ ($r\approx 10000$) and in WP$_2$($r\approx 5000$). 

The most plausible explanation is to assume that disorder comes with a variety of length scales. There is a broad distribution of the effective size of the scattering centers. The mean-free-path according to residual resistivity is long, because point-like defects (such as extrinsic atoms) do not efficiently scatter a carrier whose wavelength extends over 10 interatomic distances. The mean-free-path according to quantum oscillations is short, because such defects can affect the phase of the travelling electron. They are therefore capable of broadening Landau levels.

This interpretation would also explain the equality of Dingle mobilities in contrast to the difference in residual resistivities. The impurity content of all samples is expected to be identical, because they were grown from an identical melt, but this is not the case of dislocation density and other extended scattering centers, which can be removed by heat treatment.

The amplitude of the magnetoresistance is set by $\mu_0$ extracted from residual resistivity and not by $\mu_D$. The cleaner the sample, the larger its magnetoresistance (see Supplementary Figure \ref{fig_sup_MR}).

\subsection{Supplementary Note 4 : Low field \& low temperature recovery of the Wiedemann-Franz law}

Supplementary Figure \ref{fig_sup_WF_B} shows the thermal conductivity plotted as $\kappa/T$ as a function of $T^2$ in sample S4 in the low temperature region (where we showed the WF law to be satisfied in the main text) for three different magnetic fields. The arrows point to the value of $L_0/\rho_B$. We observe that the arrow and y-axis intercept of the linear fit match for the three magnetic fields : the WFL is recovered under the effect of these three fields. Furthermore, the slope of the linear fit to $\kappa/T(T^2)$ remains similar for the different fields. This implies that the magnetic field does not affect the lattice thermal conductivity.

\subsection{Supplementary Note 5 : Thermal conductivity and the third Onsager coefficient}
What we have measured is the thermal conductivity measured in absence of charge current. It is to be distinguished from the thermal conductivity measured in absence of electric field, which is a pure diagonal Onsager coefficient~\cite{Behnia2015b}. However, in our case, the distinction is totally negligible. The heat current density, $J^{Q}$ and the particle flow density, $J^{N}$ are Onsager fluxes responding to Onsager forces : $\nabla\frac{1}{T}$ and $\frac{1}{T}\nabla\mu$ in Supplementary Eq.(\ref{onsager1},\ref{onsager2}).

\begin{equation}
- J^{N}  = L_{11}\frac{1}{T}\nabla\mu + L_{12}\nabla\frac{1}{T}	
\label{onsager1}
\end{equation}
  \begin{equation}
J^{Q}  = L_{12}\frac{1}{T}\nabla\mu + L_{22}\nabla\frac{1}{T}		
\label{onsager2}
\end{equation}

The thermal conductivity, $\kappa$, in absence of charge current ($J^{e}=0$) and the one, $\kappa'$ in absence of potential gradient ($\nabla\mu=0$ ) are to be distinguished. The latter is inversely proportional to the Onsager coefficient $L_{22}$ as shown in Supplementary Eq.(\ref{onsager3})
\begin{equation}
\kappa' = \frac{1}{T^{2}} L_{22}	
\label{onsager3}
\end{equation}
The former is a combination of all three Onsager coefficients and its magnitude is given by Supplementary Eq.(\ref{onsager4}):
\begin{equation}
\kappa = \kappa' (1- \frac{S^{2}\sigma T}{\kappa})= \kappa' (1- \frac{S^{2}}{L})	
\label{onsager4}
\end{equation}

In our case, since $S < 5 \times 10^{-6}$V/K and $L\sim L_0 = 2.45 \times 10^{-8}$ V$^2$/K$^2$, one has $\frac{S^{2}}{L} < 0.001$, implying a negligible difference. 

\subsection{Supplementary Note 6 : Ambipolar Thermal Conductivity}

The electronic thermal conductivity of a semi-metal includes monopolar contributions from both electrons ($\kappa_e$) and holes ($\kappa_h$) as well as an ambipolar one associated with electron-hole pairs ($\kappa_{eh}$). This last contribution is negligible in Sb at $T \ll T_F$.\\
Heremans \textit{et al.} showed that the ambipolar contribution to thermal conductivity $\kappa_{eh}$ can be written as Supplementary Eq.(\ref{eq_amb_1}) \cite{heremans1977}. $\sigma_e$ and $\sigma_h$ are respectively the partial electrical conductivities associated with electrons and holes while $E_{F,e}$ and $E_{F,h}$ are the Fermi energies respectively associated with electrons and holes.
\begin{equation}
    \kappa_{eh}= (\frac{\pi^2k_B}{3e})^2T(\frac{\sigma_e\sigma_h}{\sigma_h+\sigma_e})(\frac{k_BT}{E_{f,h}}+\frac{k_BT}{E_{f,e}})^2
    \label{eq_amb_1}
\end{equation}
In the temperature range of interest of the present study, $T<10$K, the Fermi energy of holes and electrons in Sb (featured in Supplementary Table \ref{tab_semimetals}) leads to $(k_BT/E_{F,i})^2\approx10^{-4}$. This implies, at best, an ambipolar correction to the Lorenz number $L_{eh}=5.10^{-4}L_0$ at $T=10$K. Such a correction falls within the experimental error bars of this study and is consequently neglected in our discussion. The small magnitude of the Seebeck coefficient confirms this conclusion.

\subsection{Supplementary Note 7 : Viscosity, thermal conductivity and quasi-particle lifetime in Fermi liquids}

Abrikosov and Khatalnikov~\cite{Abrikosov_1959} in their 1959 seminal paper calculated the viscosity of a Fermi liquid given in Supplementary Eq.(\ref{eq_vis_1}) :
\begin{equation}
   \eta T^2=\frac{64}{45}\frac{\hbar^3p_F^5}{m^{*4}} <W_\eta>
    \label{eq_vis_1}
\end{equation}
Here $<W_\eta>$ is a temperature-independent parameter representing the angular average of scattering amplitude for viscosity, $\eta$, expected to decrease with warming as $T^{-2}$. The same collisions lead to a thermal conductivity expressed as in Supplementary Eq.(\ref{eq_vis_2}).
\begin{equation}
   \kappa T=\frac{8\pi^2}{3}\frac{\hbar^3p_F^3}{m^{*4}} <W_\kappa>
    \label{eq_vis_2}
\end{equation}

$<W_\kappa>$, like $<W_\eta>$, is neither dimensionless nor universal. The amplitude of both depends on the strength and the anisotropy of interaction and, in the case of $^3$He, strongly depends on the spin components of the overlapping wave-functions. Numerous experiments confirmed that $\eta \propto T^{-2}$~\cite{black1971,bertinat1974,alvesalo1975} and $\kappa \propto T^{-1}$~\cite{abel1967,greywall1984}. In the case of thermal conductivity, the most elaborate set of measurements performed by Greywall~\cite{greywall1984} found  that at zero pressure,  the asymptotic value for $\kappa T$ is $\kappa T|_0= 2.9 \times 10^{-4}$W.m$^{-1}$. This is about 0.6 of the theoretical value of  calculated by Brooker and Sykes ($5 \times 10^{-4}$W.m$^{-1}$)~\cite{Brooker1968}.

Calkoen and van Weert~\cite{calkoen1986} showed that in the zero temperature limit,  one  can write Supplementary Eq.(\ref{kappaT2}).
\begin{equation}
\label{kappaT2}
\kappa T|_0= \frac{5}{18\pi^3}\frac {p_F^3 v_F^2 } {A^2 }
\end{equation}
In this equation, the notation takes $\hbar=1$. In our equation 2 of the main text, in order to enhance clarity, we have introduced the dimensionless parameter $B_0$, which is simply proportional to $A^2$ as shown in Supplementary Eq.(\ref{B_0}).
\begin{equation}
\label{B_0}
B_0= \frac{9\pi^3 A^2 }{10 \hbar^2}
\end{equation}
Calkoen and van Weert~\cite{calkoen1986} found that in $^3$He, a nearly ferromagnetic liquid, the magnitude of $A$ and its variation with pressure is compatible with the Landau parameters extracted from specific heat data~\cite{greywall1983}.

The fundamental reason behind the temperature dependence of $\eta$ and $\kappa$ is the quadratic temperature dependence of the relaxation time, which can be written as Supplementary Eq.(\ref{B_0_2})~\cite{wolfle1979}: 
\begin{equation}\label{B_0_2}
\frac{\hbar}{\tau_{qp}}= \frac{(\pi k_B T)^2}{32E_F}<A>_{\theta,\phi}
\end{equation}
Here $<A>_{\theta,\phi}$ represents the angular averages of quasi-particle scattering amplitudes for transition between spin singlet and spin triplet states~\cite{wolfle1979}. In the case of $^3$He, measurements of viscosity~\cite{bertinat1974} and thermal conductivity~\cite{greywall1984} have found values for $\tau_{\kappa}T^2$ and  $\tau_{\eta}T^2$ close to each other. $\tau_{\kappa}T^2$ can be extracted from the heat capacity per volume $C_v$, using Supplementary Eq.(\ref{kappa}) :
\begin{equation}\label{kappa}
\tau_{\kappa}= 3 \frac{\kappa} {C_v v_F^2} 
\end{equation} 
As in the case of $^3$He, we have used the electronic specific heat of Sb  ($\gamma=0.105$mJ.mol$^{-1}$.K$^{-2}$)~\cite{mcCollum1967} and the average Fermi velocity to calculate $\tau_{\kappa}T^2$ in Sb.


\section{REFERENCES}

\begin{thebibliography}{48}%
\makeatletter
\providecommand \@ifxundefined [1]{%
 \@ifx{#1\undefined}
}%
\providecommand \@ifnum [1]{%
 \ifnum #1\expandafter \@firstoftwo
 \else \expandafter \@secondoftwo
 \fi
}%
\providecommand \@ifx [1]{%
 \ifx #1\expandafter \@firstoftwo
 \else \expandafter \@secondoftwo
 \fi
}%
\providecommand \natexlab [1]{#1}%
\providecommand \enquote  [1]{``#1''}%
\providecommand \bibnamefont  [1]{#1}%
\providecommand \bibfnamefont [1]{#1}%
\providecommand \citenamefont [1]{#1}%
\providecommand \href@noop [0]{\@secondoftwo}%
\providecommand \href [0]{\begingroup \@sanitize@url \@href}%
\providecommand \@href[1]{\@@startlink{#1}\@@href}%
\providecommand \@@href[1]{\endgroup#1\@@endlink}%
\providecommand \@sanitize@url [0]{\catcode `\\12\catcode `\$12\catcode
  `\&12\catcode `\#12\catcode `\^12\catcode `\_12\catcode `\%12\relax}%
\providecommand \@@startlink[1]{}%
\providecommand \@@endlink[0]{}%
\providecommand \url  [0]{\begingroup\@sanitize@url \@url }%
\providecommand \@url [1]{\endgroup\@href {#1}{\urlprefix }}%
\providecommand \urlprefix  [0]{URL }%
\providecommand \Eprint [0]{\href }%
\providecommand \doibase [0]{https://doi.org/}%
\providecommand \selectlanguage [0]{\@gobble}%
\providecommand \bibinfo  [0]{\@secondoftwo}%
\providecommand \bibfield  [0]{\@secondoftwo}%
\providecommand \translation [1]{[#1]}%
\providecommand \BibitemOpen [0]{}%
\providecommand \bibitemStop [0]{}%
\providecommand \bibitemNoStop [0]{.\EOS\space}%
\providecommand \EOS [0]{\spacefactor3000\relax}%
\providecommand \BibitemShut  [1]{\csname bibitem#1\endcsname}%
\let\auto@bib@innerbib\@empty
\bibitem [{\citenamefont {Gurzhi}(1968)}]{gurzhi1968}%
  \BibitemOpen
  \bibfield  {author} {\bibinfo {author} {\bibfnamefont {R.~N.}\ \bibnamefont
  {Gurzhi}},\ }\bibfield  {title} {\bibinfo {title} {Hydrodynamic effects in
  solids at low temperature},\ }\href@noop {} {\bibfield  {journal} {\bibinfo
  {journal} {Soviet Physics Uspekhi}\ }\textbf {\bibinfo {volume} {11}},\
  \bibinfo {pages} {255} (\bibinfo {year} {1968})}\BibitemShut {NoStop}%
\bibitem [{\citenamefont {Hartnoll}(2015)}]{hartnoll2015}%
  \BibitemOpen
  \bibfield  {author} {\bibinfo {author} {\bibfnamefont {S.~A.}\ \bibnamefont
  {Hartnoll}},\ }\bibfield  {title} {\bibinfo {title} {Theory of universal
  incoherent metallic transport},\ }\href@noop {} {\bibfield  {journal}
  {\bibinfo  {journal} {Nature Physics}\ }\textbf {\bibinfo {volume} {11}},\
  \bibinfo {pages} {54} (\bibinfo {year} {2015})}\BibitemShut {NoStop}%
\bibitem [{\citenamefont {Principi}\ and\ \citenamefont
  {Vignale}(2015)}]{principi2015}%
  \BibitemOpen
  \bibfield  {author} {\bibinfo {author} {\bibfnamefont {A.}~\bibnamefont
  {Principi}}\ and\ \bibinfo {author} {\bibfnamefont {G.}~\bibnamefont
  {Vignale}},\ }\bibfield  {title} {\bibinfo {title} {Violation of the
  {W}iedemann-{F}ranz law in hydrodynamic electron liquids},\ }\href@noop {}
  {\bibfield  {journal} {\bibinfo  {journal} {Physical Review Letters}\
  }\textbf {\bibinfo {volume} {115}},\ \bibinfo {pages} {056603} (\bibinfo
  {year} {2015})}\BibitemShut {NoStop}%
\bibitem [{\citenamefont {Scaffidi}\ \emph {et~al.}(2017)\citenamefont
  {Scaffidi}, \citenamefont {Nandi}, \citenamefont {Schmidt}, \citenamefont
  {Mackenzie},\ and\ \citenamefont {Moore}}]{scaffidi2017}%
  \BibitemOpen
  \bibfield  {author} {\bibinfo {author} {\bibfnamefont {T.}~\bibnamefont
  {Scaffidi}}, \bibinfo {author} {\bibfnamefont {N.}~\bibnamefont {Nandi}},
  \bibinfo {author} {\bibfnamefont {B.}~\bibnamefont {Schmidt}}, \bibinfo
  {author} {\bibfnamefont {A.~P.}\ \bibnamefont {Mackenzie}},\ and\ \bibinfo
  {author} {\bibfnamefont {J.~E.}\ \bibnamefont {Moore}},\ }\bibfield  {title}
  {\bibinfo {title} {Hydrodynamic electron flow and {H}all viscosity},\
  }\href@noop {} {\bibfield  {journal} {\bibinfo  {journal} {Physical Review
  Letters}\ }\textbf {\bibinfo {volume} {118}},\ \bibinfo {pages} {226601}
  (\bibinfo {year} {2017})}\BibitemShut {NoStop}%
\bibitem [{\citenamefont {de~Jong}\ and\ \citenamefont
  {Molenkamp}(1995)}]{molenkamp1994}%
  \BibitemOpen
  \bibfield  {author} {\bibinfo {author} {\bibfnamefont {M.~J.~M.}\
  \bibnamefont {de~Jong}}\ and\ \bibinfo {author} {\bibfnamefont {L.~W.}\
  \bibnamefont {Molenkamp}},\ }\bibfield  {title} {\bibinfo {title}
  {Hydrodynamic electron flow in high-mobility wires},\ }\href@noop {}
  {\bibfield  {journal} {\bibinfo  {journal} {Physical Review B}\ }\textbf
  {\bibinfo {volume} {51}},\ \bibinfo {pages} {13389} (\bibinfo {year}
  {1995})}\BibitemShut {NoStop}%
\bibitem [{\citenamefont {Moll}\ \emph {et~al.}(2016)\citenamefont {Moll},
  \citenamefont {Kushwaha}, \citenamefont {Nandi}, \citenamefont {Schmidt},\
  and\ \citenamefont {Mackenzie}}]{moll2016}%
  \BibitemOpen
  \bibfield  {author} {\bibinfo {author} {\bibfnamefont {P.~J.~W.}\
  \bibnamefont {Moll}}, \bibinfo {author} {\bibfnamefont {P.}~\bibnamefont
  {Kushwaha}}, \bibinfo {author} {\bibfnamefont {N.}~\bibnamefont {Nandi}},
  \bibinfo {author} {\bibfnamefont {B.}~\bibnamefont {Schmidt}},\ and\ \bibinfo
  {author} {\bibfnamefont {A.~P.}\ \bibnamefont {Mackenzie}},\ }\bibfield
  {title} {\bibinfo {title} {Evidence for hydrodynamic electron flow in
  {PdCoO$_2$}},\ }\href@noop {} {\bibfield  {journal} {\bibinfo  {journal}
  {Science}\ }\textbf {\bibinfo {volume} {351}},\ \bibinfo {pages} {1061}
  (\bibinfo {year} {2016})}\BibitemShut {NoStop}%
\bibitem [{\citenamefont {Crossno}\ \emph {et~al.}(2016)\citenamefont
  {Crossno}, \citenamefont {Shi}, \citenamefont {Wang}, \citenamefont {Liu},
  \citenamefont {Harzheim}, \citenamefont {Lucas}, \citenamefont {Sachdev},
  \citenamefont {Kim}, \citenamefont {Taniguchi}, \citenamefont {Watanabe}
  \emph {et~al.}}]{crossno2016}%
  \BibitemOpen
  \bibfield  {author} {\bibinfo {author} {\bibfnamefont {J.}~\bibnamefont
  {Crossno}}, \bibinfo {author} {\bibfnamefont {J.~K.}\ \bibnamefont {Shi}},
  \bibinfo {author} {\bibfnamefont {K.}~\bibnamefont {Wang}}, \bibinfo {author}
  {\bibfnamefont {X.}~\bibnamefont {Liu}}, \bibinfo {author} {\bibfnamefont
  {A.}~\bibnamefont {Harzheim}}, \bibinfo {author} {\bibfnamefont
  {A.}~\bibnamefont {Lucas}}, \bibinfo {author} {\bibfnamefont
  {S.}~\bibnamefont {Sachdev}}, \bibinfo {author} {\bibfnamefont
  {P.}~\bibnamefont {Kim}}, \bibinfo {author} {\bibfnamefont {T.}~\bibnamefont
  {Taniguchi}}, \bibinfo {author} {\bibfnamefont {K.}~\bibnamefont {Watanabe}},
  \emph {et~al.},\ }\bibfield  {title} {\bibinfo {title} {Observation of the
  dirac fluid and the breakdown of the {W}iedemann-{F}ranz law in graphene},\
  }\href@noop {} {\bibfield  {journal} {\bibinfo  {journal} {Science}\ }\textbf
  {\bibinfo {volume} {351}},\ \bibinfo {pages} {1058} (\bibinfo {year}
  {2016})}\BibitemShut {NoStop}%
\bibitem [{\citenamefont {Bandurin}\ \emph {et~al.}(2016)\citenamefont
  {Bandurin}, \citenamefont {Torre}, \citenamefont {Kumar}, \citenamefont
  {Shalom}, \citenamefont {Tomadin}, \citenamefont {Principi}, \citenamefont
  {Auton}, \citenamefont {Khestanova}, \citenamefont {Novoselov}, \citenamefont
  {Grigorieva} \emph {et~al.}}]{bandurin2016}%
  \BibitemOpen
  \bibfield  {author} {\bibinfo {author} {\bibfnamefont {D.}~\bibnamefont
  {Bandurin}}, \bibinfo {author} {\bibfnamefont {I.}~\bibnamefont {Torre}},
  \bibinfo {author} {\bibfnamefont {R.~K.}\ \bibnamefont {Kumar}}, \bibinfo
  {author} {\bibfnamefont {M.~B.}\ \bibnamefont {Shalom}}, \bibinfo {author}
  {\bibfnamefont {A.}~\bibnamefont {Tomadin}}, \bibinfo {author} {\bibfnamefont
  {A.}~\bibnamefont {Principi}}, \bibinfo {author} {\bibfnamefont
  {G.}~\bibnamefont {Auton}}, \bibinfo {author} {\bibfnamefont
  {E.}~\bibnamefont {Khestanova}}, \bibinfo {author} {\bibfnamefont
  {K.}~\bibnamefont {Novoselov}}, \bibinfo {author} {\bibfnamefont
  {I.}~\bibnamefont {Grigorieva}}, \emph {et~al.},\ }\bibfield  {title}
  {\bibinfo {title} {Negative local resistance caused by viscous electron
  backflow in graphene},\ }\href@noop {} {\bibfield  {journal} {\bibinfo
  {journal} {Science}\ }\textbf {\bibinfo {volume} {351}},\ \bibinfo {pages}
  {1055} (\bibinfo {year} {2016})}\BibitemShut {NoStop}%
\bibitem [{\citenamefont {Gooth}\ \emph {et~al.}(2018)\citenamefont {Gooth},
  \citenamefont {Menges}, \citenamefont {Kumar}, \citenamefont
  {S{\"{u}}$\beta$}, \citenamefont {Shekhar}, \citenamefont {Sun},
  \citenamefont {Drechsler}, \citenamefont {Zierold}, \citenamefont {Felser},\
  and\ \citenamefont {Gotsmann}}]{gooth2018}%
  \BibitemOpen
  \bibfield  {author} {\bibinfo {author} {\bibfnamefont {J.}~\bibnamefont
  {Gooth}}, \bibinfo {author} {\bibfnamefont {F.}~\bibnamefont {Menges}},
  \bibinfo {author} {\bibfnamefont {N.}~\bibnamefont {Kumar}}, \bibinfo
  {author} {\bibfnamefont {V.}~\bibnamefont {S{\"{u}}$\beta$}}, \bibinfo
  {author} {\bibfnamefont {C.}~\bibnamefont {Shekhar}}, \bibinfo {author}
  {\bibfnamefont {Y.}~\bibnamefont {Sun}}, \bibinfo {author} {\bibfnamefont
  {U.}~\bibnamefont {Drechsler}}, \bibinfo {author} {\bibfnamefont
  {R.}~\bibnamefont {Zierold}}, \bibinfo {author} {\bibfnamefont
  {C.}~\bibnamefont {Felser}},\ and\ \bibinfo {author} {\bibfnamefont
  {B.}~\bibnamefont {Gotsmann}},\ }\bibfield  {title} {\bibinfo {title}
  {{Thermal and electrical signatures of a hydrodynamic electron fluid in
  tungsten diphosphide}},\ }\href@noop {} {\bibfield  {journal} {\bibinfo
  {journal} {Nature Communications}\ }\textbf {\bibinfo {volume} {9}},\
  \bibinfo {pages} {4093} (\bibinfo {year} {2018})}\BibitemShut {NoStop}%
\bibitem [{\citenamefont {Sulpizio}\ \emph {et~al.}(2019)\citenamefont
  {Sulpizio}, \citenamefont {Ella}, \citenamefont {Rozen}, \citenamefont
  {Birkbeck}, \citenamefont {Perello}, \citenamefont {Dutta}, \citenamefont
  {Ben-Shalom}, \citenamefont {Taniguchi}, \citenamefont {Watanabe},
  \citenamefont {Holder} \emph {et~al.}}]{sulpizio2019}%
  \BibitemOpen
  \bibfield  {author} {\bibinfo {author} {\bibfnamefont {J.~A.}\ \bibnamefont
  {Sulpizio}}, \bibinfo {author} {\bibfnamefont {L.}~\bibnamefont {Ella}},
  \bibinfo {author} {\bibfnamefont {A.}~\bibnamefont {Rozen}}, \bibinfo
  {author} {\bibfnamefont {J.}~\bibnamefont {Birkbeck}}, \bibinfo {author}
  {\bibfnamefont {D.~J.}\ \bibnamefont {Perello}}, \bibinfo {author}
  {\bibfnamefont {D.}~\bibnamefont {Dutta}}, \bibinfo {author} {\bibfnamefont
  {M.}~\bibnamefont {Ben-Shalom}}, \bibinfo {author} {\bibfnamefont
  {T.}~\bibnamefont {Taniguchi}}, \bibinfo {author} {\bibfnamefont
  {K.}~\bibnamefont {Watanabe}}, \bibinfo {author} {\bibfnamefont
  {T.}~\bibnamefont {Holder}}, \emph {et~al.},\ }\bibfield  {title} {\bibinfo
  {title} {Visualizing Poiseuille flow of hydrodynamic electrons},\ }\href@noop
  {} {\bibfield  {journal} {\bibinfo  {journal} {Nature}\ }\textbf {\bibinfo
  {volume} {576}},\ \bibinfo {pages} {75} (\bibinfo {year} {2019})}\BibitemShut
  {NoStop}%
\bibitem [{\citenamefont {Abrikosov}\ and\ \citenamefont
  {Khalatnikov}(1959)}]{Abrikosov_1959}%
  \BibitemOpen
  \bibfield  {author} {\bibinfo {author} {\bibfnamefont {A.~A.}\ \bibnamefont
  {Abrikosov}}\ and\ \bibinfo {author} {\bibfnamefont {I.~M.}\ \bibnamefont
  {Khalatnikov}},\ }\bibfield  {title} {\bibinfo {title} {The theory of a
  {F}ermi liquid (the properties of liquid {$^3$He} at low temperatures)},\
  }\href@noop {} {\bibfield  {journal} {\bibinfo  {journal} {Reports on
  Progress in Physics}\ }\textbf {\bibinfo {volume} {22}},\ \bibinfo {pages}
  {329} (\bibinfo {year} {1959})}\BibitemShut {NoStop}%
\bibitem [{\citenamefont {{Nozi\`eres, P. and Pines D.}}(1966)}]{Nozieres}%
  \BibitemOpen
  \bibfield  {author} {\bibinfo {author} {\bibnamefont {{Nozi\`eres, P. and
  Pines D.}}},\ }\href@noop {} {\emph {\bibinfo {title} {The Theory of Quantum
  Liquids}}}\ (\bibinfo  {publisher} {CRC Press},\ \bibinfo {year}
  {1966})\BibitemShut {NoStop}%
\bibitem [{\citenamefont {Brooker}\ and\ \citenamefont
  {Sykes}(1968)}]{Brooker1968}%
  \BibitemOpen
  \bibfield  {author} {\bibinfo {author} {\bibfnamefont {G.~A.}\ \bibnamefont
  {Brooker}}\ and\ \bibinfo {author} {\bibfnamefont {J.}~\bibnamefont
  {Sykes}},\ }\bibfield  {title} {\bibinfo {title} {Transport properties of a
  {F}ermi liquid},\ }\href@noop {} {\bibfield  {journal} {\bibinfo  {journal}
  {Physical Review Letters}\ }\textbf {\bibinfo {volume} {21}},\ \bibinfo
  {pages} {279} (\bibinfo {year} {1968})}\BibitemShut {NoStop}%
\bibitem [{\citenamefont {Wheatley}(1968)}]{Wheatley1968}%
  \BibitemOpen
  \bibfield  {author} {\bibinfo {author} {\bibfnamefont {J.~C.}\ \bibnamefont
  {Wheatley}},\ }\bibfield  {title} {\bibinfo {title} {Experimental properties
  of liquid ${\mathrm{He}}^{3}$ near the absolute zero},\ }\href@noop {}
  {\bibfield  {journal} {\bibinfo  {journal} {Physical Review}\ }\textbf
  {\bibinfo {volume} {165}},\ \bibinfo {pages} {304} (\bibinfo {year}
  {1968})}\BibitemShut {NoStop}%
\bibitem [{\citenamefont {Greywall}(1984)}]{greywall1984}%
  \BibitemOpen
  \bibfield  {author} {\bibinfo {author} {\bibfnamefont {D.~S.}\ \bibnamefont
  {Greywall}},\ }\bibfield  {title} {\bibinfo {title} {Thermal conductivity of
  normal liquid {$^3$He}},\ }\href@noop {} {\bibfield  {journal} {\bibinfo
  {journal} {Physical Review B}\ }\textbf {\bibinfo {volume} {29}},\ \bibinfo
  {pages} {4933} (\bibinfo {year} {1984})}\BibitemShut {NoStop}%
\bibitem [{\citenamefont {Bertinat}\ \emph {et~al.}(1974)\citenamefont
  {Bertinat}, \citenamefont {Betts}, \citenamefont {Brewer},\ and\
  \citenamefont {Butterworth}}]{bertinat1974}%
  \BibitemOpen
  \bibfield  {author} {\bibinfo {author} {\bibfnamefont {M.~P.}\ \bibnamefont
  {Bertinat}}, \bibinfo {author} {\bibfnamefont {D.~S.}\ \bibnamefont {Betts}},
  \bibinfo {author} {\bibfnamefont {D.~F.}\ \bibnamefont {Brewer}},\ and\
  \bibinfo {author} {\bibfnamefont {G.~J.}\ \bibnamefont {Butterworth}},\
  }\bibfield  {title} {\bibinfo {title} {Damping of torsional oscillations of a
  quartz crystal cylinder in liquid helium at low temperatures. {I}. viscosity
  of pure {$^3$H}e},\ }\href@noop {} {\bibfield  {journal} {\bibinfo  {journal}
  {Journal of Low Temperature Physics}\ }\textbf {\bibinfo {volume} {16}},\
  \bibinfo {pages} {479} (\bibinfo {year} {1974})}\BibitemShut {NoStop}%
\bibitem [{\citenamefont {Alvesalo}\ \emph {et~al.}(1975)\citenamefont
  {Alvesalo}, \citenamefont {Collan}, \citenamefont {Loponen}, \citenamefont
  {Lounasmaa},\ and\ \citenamefont {Veuro}}]{alvesalo1975}%
  \BibitemOpen
  \bibfield  {author} {\bibinfo {author} {\bibfnamefont {T.~A.}\ \bibnamefont
  {Alvesalo}}, \bibinfo {author} {\bibfnamefont {H.~K.}\ \bibnamefont
  {Collan}}, \bibinfo {author} {\bibfnamefont {M.~T.}\ \bibnamefont {Loponen}},
  \bibinfo {author} {\bibfnamefont {O.~V.}\ \bibnamefont {Lounasmaa}},\ and\
  \bibinfo {author} {\bibfnamefont {M.~C.}\ \bibnamefont {Veuro}},\ }\bibfield
  {title} {\bibinfo {title} {The viscosity and some related properties of
  liquid {$^3$H}e at the melting curve between 1 and 100 mK},\ }\href@noop {}
  {\bibfield  {journal} {\bibinfo  {journal} {Journal of Low Temperature
  Physics}\ }\textbf {\bibinfo {volume} {19}},\ \bibinfo {pages} {1} (\bibinfo
  {year} {1975})}\BibitemShut {NoStop}%
\bibitem [{\citenamefont {Ziman}(1972)}]{ziman1972}%
  \BibitemOpen
  \bibfield  {author} {\bibinfo {author} {\bibfnamefont {J.}~\bibnamefont
  {Ziman}},\ }\href@noop {} {\emph {\bibinfo {title} {Principles of the Theory
  of Solids}}}\ (\bibinfo  {publisher} {Cambridge University Press},\ \bibinfo
  {year} {1972})\BibitemShut {NoStop}%
\bibitem [{\citenamefont {Wagner}\ \emph {et~al.}(1971)\citenamefont {Wagner},
  \citenamefont {Garland},\ and\ \citenamefont {Bowers}}]{wagner1971}%
  \BibitemOpen
  \bibfield  {author} {\bibinfo {author} {\bibfnamefont {D.~K.}\ \bibnamefont
  {Wagner}}, \bibinfo {author} {\bibfnamefont {J.~C.}\ \bibnamefont
  {Garland}},\ and\ \bibinfo {author} {\bibfnamefont {R.}~\bibnamefont
  {Bowers}},\ }\bibfield  {title} {\bibinfo {title} {Low-temperature electrical
  and thermal resistivities of tungsten},\ }\href@noop {} {\bibfield  {journal}
  {\bibinfo  {journal} {Physical Review B}\ }\textbf {\bibinfo {volume} {3}},\
  \bibinfo {pages} {3141} (\bibinfo {year} {1971})}\BibitemShut {NoStop}%
\bibitem [{\citenamefont {Paglione}\ \emph {et~al.}(2005)\citenamefont
  {Paglione}, \citenamefont {Tanatar}, \citenamefont {Hawthorn}, \citenamefont
  {Hill}, \citenamefont {Ronning}, \citenamefont {Sutherland}, \citenamefont
  {Taillefer}, \citenamefont {Petrovic},\ and\ \citenamefont
  {Canfield}}]{paglione2005}%
  \BibitemOpen
  \bibfield  {author} {\bibinfo {author} {\bibfnamefont {J.}~\bibnamefont
  {Paglione}}, \bibinfo {author} {\bibfnamefont {M.~A.}\ \bibnamefont
  {Tanatar}}, \bibinfo {author} {\bibfnamefont {D.~G.}\ \bibnamefont
  {Hawthorn}}, \bibinfo {author} {\bibfnamefont {R.~W.}\ \bibnamefont {Hill}},
  \bibinfo {author} {\bibfnamefont {F.}~\bibnamefont {Ronning}}, \bibinfo
  {author} {\bibfnamefont {M.}~\bibnamefont {Sutherland}}, \bibinfo {author}
  {\bibfnamefont {L.}~\bibnamefont {Taillefer}}, \bibinfo {author}
  {\bibfnamefont {C.}~\bibnamefont {Petrovic}},\ and\ \bibinfo {author}
  {\bibfnamefont {P.~C.}\ \bibnamefont {Canfield}},\ }\bibfield  {title}
  {\bibinfo {title} {Heat transport as a probe of electron scattering by spin
  fluctuations: the case of antiferromagnetic {CeRhIn$_5$}},\ }\href@noop {}
  {\bibfield  {journal} {\bibinfo  {journal} {Physical Review Letters}\
  }\textbf {\bibinfo {volume} {94}},\ \bibinfo {pages} {216602} (\bibinfo
  {year} {2005})}\BibitemShut {NoStop}%
\bibitem [{\citenamefont {Jaoui}\ \emph {et~al.}(2018)\citenamefont {Jaoui},
  \citenamefont {Fauqu{\'e}}, \citenamefont {Rischau}, \citenamefont {Subedi},
  \citenamefont {Fu}, \citenamefont {Gooth}, \citenamefont {Kumar},
  \citenamefont {S{\"u}{\ss}}, \citenamefont {Maslov}, \citenamefont {Felser},\
  and\ \citenamefont {Behnia}}]{jaoui2018}%
  \BibitemOpen
  \bibfield  {author} {\bibinfo {author} {\bibfnamefont {A.}~\bibnamefont
  {Jaoui}}, \bibinfo {author} {\bibfnamefont {B.}~\bibnamefont {Fauqu{\'e}}},
  \bibinfo {author} {\bibfnamefont {C.~W.}\ \bibnamefont {Rischau}}, \bibinfo
  {author} {\bibfnamefont {A.}~\bibnamefont {Subedi}}, \bibinfo {author}
  {\bibfnamefont {C.}~\bibnamefont {Fu}}, \bibinfo {author} {\bibfnamefont
  {J.}~\bibnamefont {Gooth}}, \bibinfo {author} {\bibfnamefont
  {N.}~\bibnamefont {Kumar}}, \bibinfo {author} {\bibfnamefont
  {V.}~\bibnamefont {S{\"u}{\ss}}}, \bibinfo {author} {\bibfnamefont {D.~L.}\
  \bibnamefont {Maslov}}, \bibinfo {author} {\bibfnamefont {C.}~\bibnamefont
  {Felser}},\ and\ \bibinfo {author} {\bibfnamefont {K.}~\bibnamefont
  {Behnia}},\ }\bibfield  {title} {\bibinfo {title} {Departure from the
  {W}iedemann--{F}ranz law in {WP$_2$} driven by mismatch in {T-}square
  resistivity prefactors},\ }\href@noop {} {\bibfield  {journal} {\bibinfo
  {journal} {npj Quantum Materials}\ }\textbf {\bibinfo {volume} {3}},\
  \bibinfo {pages} {64} (\bibinfo {year} {2018})}\BibitemShut {NoStop}%
\bibitem [{\citenamefont {Li}\ and\ \citenamefont {Maslov}(2018)}]{li2018}%
  \BibitemOpen
  \bibfield  {author} {\bibinfo {author} {\bibfnamefont {S.}~\bibnamefont
  {Li}}\ and\ \bibinfo {author} {\bibfnamefont {D.~L.}\ \bibnamefont
  {Maslov}},\ }\bibfield  {title} {\bibinfo {title} {Lorentz ratio of a
  compensated metal},\ }\href@noop {} {\bibfield  {journal} {\bibinfo
  {journal} {Physical Review B}\ }\textbf {\bibinfo {volume} {98}},\ \bibinfo
  {pages} {245134} (\bibinfo {year} {2018})}\BibitemShut {NoStop}%
\bibitem [{\citenamefont {Fauqu\'e}\ \emph {et~al.}(2018)\citenamefont
  {Fauqu\'e}, \citenamefont {Yang}, \citenamefont {Tabis}, \citenamefont
  {Shen}, \citenamefont {Zhu}, \citenamefont {Proust}, \citenamefont {Fuseya},\
  and\ \citenamefont {Behnia}}]{fauque2018}%
  \BibitemOpen
  \bibfield  {author} {\bibinfo {author} {\bibfnamefont {B.}~\bibnamefont
  {Fauqu\'e}}, \bibinfo {author} {\bibfnamefont {X.}~\bibnamefont {Yang}},
  \bibinfo {author} {\bibfnamefont {W.}~\bibnamefont {Tabis}}, \bibinfo
  {author} {\bibfnamefont {M.}~\bibnamefont {Shen}}, \bibinfo {author}
  {\bibfnamefont {Z.}~\bibnamefont {Zhu}}, \bibinfo {author} {\bibfnamefont
  {C.}~\bibnamefont {Proust}}, \bibinfo {author} {\bibfnamefont
  {Y.}~\bibnamefont {Fuseya}},\ and\ \bibinfo {author} {\bibfnamefont
  {K.}~\bibnamefont {Behnia}},\ }\bibfield  {title} {\bibinfo {title}
  {Magnetoresistance of semimetals: the case of antimony},\ }\href@noop {}
  {\bibfield  {journal} {\bibinfo  {journal} {Physical Review Materials}\
  }\textbf {\bibinfo {volume} {2}},\ \bibinfo {pages} {114201} (\bibinfo {year}
  {2018})}\BibitemShut {NoStop}%
\bibitem [{\citenamefont {{Bogod}}\ and\ \citenamefont
  {{Krasovtski{\v{i}}}}(1973)}]{Bogod1973}%
  \BibitemOpen
  \bibfield  {author} {\bibinfo {author} {\bibfnamefont {Y.~A.}\ \bibnamefont
  {{Bogod}}}\ and\ \bibinfo {author} {\bibfnamefont {V.~B.}\ \bibnamefont
  {{Krasovtski{\v{i}}}}},\ }\bibfield  {title} {\bibinfo {title}
  {{Galvanomagnetic properties of antimony at low temperatures. Size effect,
  role of surface and shape effects}},\ }\href@noop {} {\bibfield  {journal}
  {\bibinfo  {journal} {Soviet Journal of Experimental and Theoretical
  Physics}\ }\textbf {\bibinfo {volume} {36}},\ \bibinfo {pages} {544}
  (\bibinfo {year} {1973})}\BibitemShut {NoStop}%
\bibitem [{\citenamefont {Herrod}\ \emph {et~al.}(1971)\citenamefont {Herrod},
  \citenamefont {Gage},\ and\ \citenamefont {Goodrich}}]{herrod1971}%
  \BibitemOpen
  \bibfield  {author} {\bibinfo {author} {\bibfnamefont {R.}~\bibnamefont
  {Herrod}}, \bibinfo {author} {\bibfnamefont {C.}~\bibnamefont {Gage}},\ and\
  \bibinfo {author} {\bibfnamefont {R.}~\bibnamefont {Goodrich}},\ }\bibfield
  {title} {\bibinfo {title} {Fermi surface of antimony: radio-frequency size
  zffect},\ }\href@noop {} {\bibfield  {journal} {\bibinfo  {journal} {Physical
  Review B}\ }\textbf {\bibinfo {volume} {4}},\ \bibinfo {pages} {1033}
  (\bibinfo {year} {1971})}\BibitemShut {NoStop}%
\bibitem [{\citenamefont {Issi}(1979)}]{issi1979}%
  \BibitemOpen
  \bibfield  {author} {\bibinfo {author} {\bibfnamefont {J.}~\bibnamefont
  {Issi}},\ }\bibfield  {title} {\bibinfo {title} {Low temperature transport
  properties of the group V semimetals},\ }\href@noop {} {\bibfield  {journal}
  {\bibinfo  {journal} {Australian Journal of Physics}\ }\textbf {\bibinfo
  {volume} {32}},\ \bibinfo {pages} {585} (\bibinfo {year} {1979})}\BibitemShut
  {NoStop}%
\bibitem [{\citenamefont {Gonze}\ \emph {et~al.}(1990)\citenamefont {Gonze},
  \citenamefont {Michenaud},\ and\ \citenamefont {Vigneron}}]{Gonze1990}%
  \BibitemOpen
  \bibfield  {author} {\bibinfo {author} {\bibfnamefont {X.}~\bibnamefont
  {Gonze}}, \bibinfo {author} {\bibfnamefont {J.-P.}\ \bibnamefont
  {Michenaud}},\ and\ \bibinfo {author} {\bibfnamefont {J.-P.}\ \bibnamefont
  {Vigneron}},\ }\bibfield  {title} {\bibinfo {title} {First-principles study
  of As, Sb, and Bi electronic properties},\ }\href@noop {} {\bibfield
  {journal} {\bibinfo  {journal} {Physical Review B}\ }\textbf {\bibinfo
  {volume} {41}},\ \bibinfo {pages} {11827} (\bibinfo {year}
  {1990})}\BibitemShut {NoStop}%
\bibitem [{\citenamefont {Liu}\ and\ \citenamefont {Allen}(1995)}]{liu1995}%
  \BibitemOpen
  \bibfield  {author} {\bibinfo {author} {\bibfnamefont {Y.}~\bibnamefont
  {Liu}}\ and\ \bibinfo {author} {\bibfnamefont {R.~E.}\ \bibnamefont
  {Allen}},\ }\bibfield  {title} {\bibinfo {title} {Electronic structure of the
  semimetals {Bi} and {Sb}},\ }\href@noop {} {\bibfield  {journal} {\bibinfo
  {journal} {Physical Review B}\ }\textbf {\bibinfo {volume} {52}},\ \bibinfo
  {pages} {1566} (\bibinfo {year} {1995})}\BibitemShut {NoStop}%
\bibitem [{\citenamefont {Behnia}(2015)}]{Behnia2015b}%
  \BibitemOpen
  \bibfield  {author} {\bibinfo {author} {\bibfnamefont {K.}~\bibnamefont
  {Behnia}},\ }\href@noop {} {\emph {\bibinfo {title} {{Fundamentals of
  Thermoelectricity}}}}\ (\bibinfo  {publisher} {Oxford University Press},\
  \bibinfo {year} {2015})\BibitemShut {NoStop}%
\bibitem [{\citenamefont {Uher}\ and\ \citenamefont
  {Goldsmid}(1974)}]{Uher1974}%
  \BibitemOpen
  \bibfield  {author} {\bibinfo {author} {\bibfnamefont {C.}~\bibnamefont
  {Uher}}\ and\ \bibinfo {author} {\bibfnamefont {H.~J.}\ \bibnamefont
  {Goldsmid}},\ }\bibfield  {title} {\bibinfo {title} {Separation of the
  electronic and lattice thermal conductivities in bismuth crystals},\
  }\href@noop {} {\bibfield  {journal} {\bibinfo  {journal} {Physica Status
  Solidi B}\ }\textbf {\bibinfo {volume} {65}},\ \bibinfo {pages} {765}
  (\bibinfo {year} {1974})}\BibitemShut {NoStop}%
\bibitem [{\citenamefont {Tsai}\ \emph {et~al.}(1978)\citenamefont {Tsai},
  \citenamefont {Waldorf}, \citenamefont {Tanaka},\ and\ \citenamefont
  {Grenier}}]{tsai1978}%
  \BibitemOpen
  \bibfield  {author} {\bibinfo {author} {\bibfnamefont {C.~L.}\ \bibnamefont
  {Tsai}}, \bibinfo {author} {\bibfnamefont {D.}~\bibnamefont {Waldorf}},
  \bibinfo {author} {\bibfnamefont {K.}~\bibnamefont {Tanaka}},\ and\ \bibinfo
  {author} {\bibfnamefont {C.~G.}\ \bibnamefont {Grenier}},\ }\bibfield
  {title} {\bibinfo {title} {Mutual drag effect in the magnetoresistivity of
  antimony},\ }\href@noop {} {\bibfield  {journal} {\bibinfo  {journal}
  {Physical Review B}\ }\textbf {\bibinfo {volume} {17}},\ \bibinfo {pages}
  {618} (\bibinfo {year} {1978})}\BibitemShut {NoStop}%
\bibitem [{\citenamefont {Bresler}\ and\ \citenamefont
  {Red'ko}(1972)}]{bresler1972}%
  \BibitemOpen
  \bibfield  {author} {\bibinfo {author} {\bibfnamefont {M.~S.}\ \bibnamefont
  {Bresler}}\ and\ \bibinfo {author} {\bibfnamefont {N.~A.}\ \bibnamefont
  {Red'ko}},\ }\bibfield  {title} {\bibinfo {title} {{Galvanomagnetic phenomena
  in antimony at low temperatures}},\ }\href@noop {} {\bibfield  {journal}
  {\bibinfo  {journal} {Soviet Journal of Experimental and Theoretical
  Physics}\ }\textbf {\bibinfo {volume} {34}},\ \bibinfo {pages} {149}
  (\bibinfo {year} {1972})}\BibitemShut {NoStop}%
\bibitem [{\citenamefont {Lussier}\ \emph {et~al.}(1994)\citenamefont
  {Lussier}, \citenamefont {Ellman},\ and\ \citenamefont
  {Taillefer}}]{lussier1994}%
  \BibitemOpen
  \bibfield  {author} {\bibinfo {author} {\bibfnamefont {B.}~\bibnamefont
  {Lussier}}, \bibinfo {author} {\bibfnamefont {B.}~\bibnamefont {Ellman}},\
  and\ \bibinfo {author} {\bibfnamefont {L.}~\bibnamefont {Taillefer}},\
  }\bibfield  {title} {\bibinfo {title} {Anisotropy of heat conduction in the
  heavy fermion superconductor {UPt$_3$}},\ }\href@noop {} {\bibfield
  {journal} {\bibinfo  {journal} {Physical Review Letters}\ }\textbf {\bibinfo
  {volume} {73}},\ \bibinfo {pages} {3294} (\bibinfo {year}
  {1994})}\BibitemShut {NoStop}%
\bibitem [{\citenamefont {Lin}\ \emph {et~al.}(2015)\citenamefont {Lin},
  \citenamefont {Fauqu{\'e}},\ and\ \citenamefont {Behnia}}]{lin2015}%
  \BibitemOpen
  \bibfield  {author} {\bibinfo {author} {\bibfnamefont {X.}~\bibnamefont
  {Lin}}, \bibinfo {author} {\bibfnamefont {B.}~\bibnamefont {Fauqu{\'e}}},\
  and\ \bibinfo {author} {\bibfnamefont {K.}~\bibnamefont {Behnia}},\
  }\bibfield  {title} {\bibinfo {title} {Scalable {$T^2$} resistivity in a
  small single-component Fermi surface},\ }\href@noop {} {\bibfield  {journal}
  {\bibinfo  {journal} {Science}\ }\textbf {\bibinfo {volume} {349}},\ \bibinfo
  {pages} {945} (\bibinfo {year} {2015})}\BibitemShut {NoStop}%
\bibitem [{\citenamefont {Wang}\ \emph {et~al.}(2020)\citenamefont {Wang},
  \citenamefont {Wu}, \citenamefont {Wang}, \citenamefont {Xu}, \citenamefont
  {Wu}, \citenamefont {Hu}, \citenamefont {Ren}, \citenamefont {Liu},
  \citenamefont {Behnia},\ and\ \citenamefont {Lin}}]{Wang2020}%
  \BibitemOpen
  \bibfield  {author} {\bibinfo {author} {\bibfnamefont {J.}~\bibnamefont
  {Wang}}, \bibinfo {author} {\bibfnamefont {J.}~\bibnamefont {Wu}}, \bibinfo
  {author} {\bibfnamefont {T.}~\bibnamefont {Wang}}, \bibinfo {author}
  {\bibfnamefont {Z.}~\bibnamefont {Xu}}, \bibinfo {author} {\bibfnamefont
  {J.}~\bibnamefont {Wu}}, \bibinfo {author} {\bibfnamefont {W.}~\bibnamefont
  {Hu}}, \bibinfo {author} {\bibfnamefont {Z.}~\bibnamefont {Ren}}, \bibinfo
  {author} {\bibfnamefont {S.}~\bibnamefont {Liu}}, \bibinfo {author}
  {\bibfnamefont {K.}~\bibnamefont {Behnia}},\ and\ \bibinfo {author}
  {\bibfnamefont {X.}~\bibnamefont {Lin}},\ }\bibfield  {title} {\bibinfo
  {title} {T-square resistivity without umklapp scattering in dilute metallic
  Bi$_2$O$_2$Se},\ }\href@noop {} {\bibfield  {journal} {\bibinfo  {journal}
  {Nature Communications}\ }\textbf {\bibinfo {volume} {11}},\ \bibinfo {pages}
  {3846} (\bibinfo {year} {2020})}\BibitemShut {NoStop}%
\bibitem [{\citenamefont {Sambles}\ and\ \citenamefont
  {Elson}(1980)}]{sambles1980}%
  \BibitemOpen
  \bibfield  {author} {\bibinfo {author} {\bibfnamefont {J.~R.}\ \bibnamefont
  {Sambles}}\ and\ \bibinfo {author} {\bibfnamefont {K.~C.}\ \bibnamefont
  {Elson}},\ }\bibfield  {title} {\bibinfo {title} {Electrical conduction in
  metal foils},\ }\href@noop {} {\bibfield  {journal} {\bibinfo  {journal}
  {Journal of Physics F: Metal Physics}\ }\textbf {\bibinfo {volume} {10}},\
  \bibinfo {pages} {1487} (\bibinfo {year} {1980})}\BibitemShut {NoStop}%
\bibitem [{\citenamefont {van~der Maas}\ \emph {et~al.}(1985)\citenamefont
  {van~der Maas}, \citenamefont {Huguenin},\ and\ \citenamefont
  {Gasparov}}]{maas1985}%
  \BibitemOpen
  \bibfield  {author} {\bibinfo {author} {\bibfnamefont {J.}~\bibnamefont
  {van~der Maas}}, \bibinfo {author} {\bibfnamefont {R.}~\bibnamefont
  {Huguenin}},\ and\ \bibinfo {author} {\bibfnamefont {V.~A.}\ \bibnamefont
  {Gasparov}},\ }\bibfield  {title} {\bibinfo {title} {Electron-electron
  scattering in tungsten},\ }\href@noop {} {\bibfield  {journal} {\bibinfo
  {journal} {Journal of Physics F: Metal Physics}\ }\textbf {\bibinfo {volume}
  {15}},\ \bibinfo {pages} {271} (\bibinfo {year} {1985})}\BibitemShut
  {NoStop}%
\bibitem [{\citenamefont {{Soffer}}(1967)}]{soffer1967}%
  \BibitemOpen
  \bibfield  {author} {\bibinfo {author} {\bibfnamefont {S.~B.}\ \bibnamefont
  {{Soffer}}},\ }\bibfield  {title} {\bibinfo {title} {{Statistical model for
  the size effect in electrical conduction}},\ }\href@noop {} {\bibfield
  {journal} {\bibinfo  {journal} {Journal of Applied Physics}\ }\textbf
  {\bibinfo {volume} {38}},\ \bibinfo {pages} {1710} (\bibinfo {year}
  {1967})}\BibitemShut {NoStop}%
\bibitem [{\citenamefont {Sambles}\ and\ \citenamefont
  {Mundy}(1983)}]{sambles1983}%
  \BibitemOpen
  \bibfield  {author} {\bibinfo {author} {\bibfnamefont {J.~R.}\ \bibnamefont
  {Sambles}}\ and\ \bibinfo {author} {\bibfnamefont {J.~N.}\ \bibnamefont
  {Mundy}},\ }\bibfield  {title} {\bibinfo {title} {A reanalysis of resistive
  size effects in tungsten},\ }\href@noop {} {\bibfield  {journal} {\bibinfo
  {journal} {Journal of Physics F: Metal Physics}\ }\textbf {\bibinfo {volume}
  {13}},\ \bibinfo {pages} {2281} (\bibinfo {year} {1983})}\BibitemShut
  {NoStop}%
\bibitem [{\citenamefont {Kiselev}\ and\ \citenamefont
  {Schmalian}(2019)}]{kiselev2019}%
  \BibitemOpen
  \bibfield  {author} {\bibinfo {author} {\bibfnamefont {E.~I.}\ \bibnamefont
  {Kiselev}}\ and\ \bibinfo {author} {\bibfnamefont {J.}~\bibnamefont
  {Schmalian}},\ }\bibfield  {title} {\bibinfo {title} {Boundary conditions of
  viscous electron flow},\ }\href@noop {} {\bibfield  {journal} {\bibinfo
  {journal} {Physical Review B}\ }\textbf {\bibinfo {volume} {99}},\ \bibinfo
  {pages} {035430} (\bibinfo {year} {2019})}\BibitemShut {NoStop}%
\bibitem [{\citenamefont {Calkoen}\ and\ \citenamefont {van
  Weert}(1986)}]{calkoen1986}%
  \BibitemOpen
  \bibfield  {author} {\bibinfo {author} {\bibfnamefont {C.~J.}\ \bibnamefont
  {Calkoen}}\ and\ \bibinfo {author} {\bibfnamefont {C.~G.}\ \bibnamefont {van
  Weert}},\ }\bibfield  {title} {\bibinfo {title} {Thermal conductivity of
  normal liquid {$^3$H}e at finite temperatures},\ }\href@noop {} {\bibfield
  {journal} {\bibinfo  {journal} {Journal of Low Temperature Physics}\ }\textbf
  {\bibinfo {volume} {64}},\ \bibinfo {pages} {429} (\bibinfo {year}
  {1986})}\BibitemShut {NoStop}%
\bibitem [{\citenamefont {Greywall}(1983)}]{greywall1983}%
  \BibitemOpen
  \bibfield  {author} {\bibinfo {author} {\bibfnamefont {D.~S.}\ \bibnamefont
  {Greywall}},\ }\bibfield  {title} {\bibinfo {title} {Specific heat of normal
  liquid $^{3}\mathrm{He}$},\ }\href@noop {} {\bibfield  {journal} {\bibinfo
  {journal} {Physical Review B}\ }\textbf {\bibinfo {volume} {27}},\ \bibinfo
  {pages} {2747} (\bibinfo {year} {1983})}\BibitemShut {NoStop}%
\bibitem [{\citenamefont {Wolfle}(1979)}]{wolfle1979}%
  \BibitemOpen
  \bibfield  {author} {\bibinfo {author} {\bibfnamefont {P.}~\bibnamefont
  {Wolfle}},\ }\bibfield  {title} {\bibinfo {title} {Low-temperature properties
  of liquid {$^3$H}e},\ }\href@noop {} {\bibfield  {journal} {\bibinfo
  {journal} {Reports on Progress in Physics}\ }\textbf {\bibinfo {volume}
  {42}},\ \bibinfo {pages} {269} (\bibinfo {year} {1979})}\BibitemShut
  {NoStop}%
\bibitem [{\citenamefont {Collaudin}\ \emph {et~al.}(2015)\citenamefont
  {Collaudin}, \citenamefont {Fauqu\'e}, \citenamefont {Fuseya}, \citenamefont
  {Kang},\ and\ \citenamefont {Behnia}}]{collaudin2015}%
  \BibitemOpen
  \bibfield  {author} {\bibinfo {author} {\bibfnamefont {A.}~\bibnamefont
  {Collaudin}}, \bibinfo {author} {\bibfnamefont {B.}~\bibnamefont {Fauqu\'e}},
  \bibinfo {author} {\bibfnamefont {Y.}~\bibnamefont {Fuseya}}, \bibinfo
  {author} {\bibfnamefont {W.}~\bibnamefont {Kang}},\ and\ \bibinfo {author}
  {\bibfnamefont {K.}~\bibnamefont {Behnia}},\ }\bibfield  {title} {\bibinfo
  {title} {Angle dependence of the orbital magnetoresistance in bismuth},\
  }\href@noop {} {\bibfield  {journal} {\bibinfo  {journal} {Physical Review
  X}\ }\textbf {\bibinfo {volume} {5}},\ \bibinfo {pages} {021022} (\bibinfo
  {year} {2015})}\BibitemShut {NoStop}%
\bibitem [{\citenamefont {Hicks}\ \emph {et~al.}(2012)\citenamefont {Hicks},
  \citenamefont {Gibbs}, \citenamefont {Mackenzie}, \citenamefont {Takatsu},
  \citenamefont {Maeno},\ and\ \citenamefont {Yelland}}]{Hicks2012}%
  \BibitemOpen
  \bibfield  {author} {\bibinfo {author} {\bibfnamefont {C.~W.}\ \bibnamefont
  {Hicks}}, \bibinfo {author} {\bibfnamefont {A.~S.}\ \bibnamefont {Gibbs}},
  \bibinfo {author} {\bibfnamefont {A.~P.}\ \bibnamefont {Mackenzie}}, \bibinfo
  {author} {\bibfnamefont {H.}~\bibnamefont {Takatsu}}, \bibinfo {author}
  {\bibfnamefont {Y.}~\bibnamefont {Maeno}},\ and\ \bibinfo {author}
  {\bibfnamefont {E.~A.}\ \bibnamefont {Yelland}},\ }\bibfield  {title}
  {\bibinfo {title} {Quantum oscillations and high carrier mobility in the
  delafossite ${\mathrm{PdCoO}}_{2}$},\ }\href@noop {} {\bibfield  {journal}
  {\bibinfo  {journal} {Physical Review Letters}\ }\textbf {\bibinfo {volume}
  {109}},\ \bibinfo {pages} {116401} (\bibinfo {year} {2012})}\BibitemShut
  {NoStop}%
\bibitem [{\citenamefont {Mackenzie}(2017)}]{Mackenzie_2017}%
  \BibitemOpen
  \bibfield  {author} {\bibinfo {author} {\bibfnamefont {A.~P.}\ \bibnamefont
  {Mackenzie}},\ }\bibfield  {title} {\bibinfo {title} {The properties of
  ultrapure delafossite metals},\ }\href@noop {} {\bibfield  {journal}
  {\bibinfo  {journal} {Reports on Progress in Physics}\ }\textbf {\bibinfo
  {volume} {80}},\ \bibinfo {pages} {032501} (\bibinfo {year}
  {2017})}\BibitemShut {NoStop}%
\bibitem [{\citenamefont {Kasahara}\ \emph {et~al.}(2007)\citenamefont
  {Kasahara}, \citenamefont {Iwasawa}, \citenamefont {Shishido}, \citenamefont
  {Shibauchi}, \citenamefont {Behnia}, \citenamefont {Haga}, \citenamefont
  {Matsuda}, \citenamefont {Onuki}, \citenamefont {Sigrist},\ and\
  \citenamefont {Matsuda}}]{Kasahara2007}%
  \BibitemOpen
  \bibfield  {author} {\bibinfo {author} {\bibfnamefont {Y.}~\bibnamefont
  {Kasahara}}, \bibinfo {author} {\bibfnamefont {T.}~\bibnamefont {Iwasawa}},
  \bibinfo {author} {\bibfnamefont {H.}~\bibnamefont {Shishido}}, \bibinfo
  {author} {\bibfnamefont {T.}~\bibnamefont {Shibauchi}}, \bibinfo {author}
  {\bibfnamefont {K.}~\bibnamefont {Behnia}}, \bibinfo {author} {\bibfnamefont
  {Y.}~\bibnamefont {Haga}}, \bibinfo {author} {\bibfnamefont {T.~D.}\
  \bibnamefont {Matsuda}}, \bibinfo {author} {\bibfnamefont {Y.}~\bibnamefont
  {Onuki}}, \bibinfo {author} {\bibfnamefont {M.}~\bibnamefont {Sigrist}},\
  and\ \bibinfo {author} {\bibfnamefont {Y.}~\bibnamefont {Matsuda}},\
  }\bibfield  {title} {\bibinfo {title} {Exotic superconducting properties in
  the electron-hole-compensated heavy-fermion ``semimetal''
  ${\mathrm{URu}}_{2}{\mathrm{Si}}_{2}$},\ }\href@noop {} {\bibfield  {journal}
  {\bibinfo  {journal} {Physical Review Letters}\ }\textbf {\bibinfo {volume}
  {99}},\ \bibinfo {pages} {116402} (\bibinfo {year} {2007})}\BibitemShut
  {NoStop}%
\bibitem [{\citenamefont {Pourret}\ \emph {et~al.}(2006)\citenamefont
  {Pourret}, \citenamefont {Behnia}, \citenamefont {Kikuchi}, \citenamefont
  {Aoki}, \citenamefont {Sugawara},\ and\ \citenamefont {Sato}}]{Pourret2006}%
  \BibitemOpen
  \bibfield  {author} {\bibinfo {author} {\bibfnamefont {A.}~\bibnamefont
  {Pourret}}, \bibinfo {author} {\bibfnamefont {K.}~\bibnamefont {Behnia}},
  \bibinfo {author} {\bibfnamefont {D.}~\bibnamefont {Kikuchi}}, \bibinfo
  {author} {\bibfnamefont {Y.}~\bibnamefont {Aoki}}, \bibinfo {author}
  {\bibfnamefont {H.}~\bibnamefont {Sugawara}},\ and\ \bibinfo {author}
  {\bibfnamefont {H.}~\bibnamefont {Sato}},\ }\bibfield  {title} {\bibinfo
  {title} {Drastic change in transport of entropy with quadrupolar ordering in
  ${\mathrm{PrFe}}_{4}{\mathrm{P}}_{12}$},\ }\href@noop {} {\bibfield
  {journal} {\bibinfo  {journal} {Physical Review Letters}\ }\textbf {\bibinfo
  {volume} {96}},\ \bibinfo {pages} {176402} (\bibinfo {year}
  {2006})}\BibitemShut {NoStop}%
\bibitem{edelman1976}
V.~S. Edelman, Electrons in bismuth, Advances in Physics \textbf{25}, 555-613 (1976).
\bibitem{zhu2015}
Z. Zhu, X. Lin, J. Liu, B. Fauqu{\'e}, Q. Tao, C. Yang, Y. Shi, and K. Behnia, Quantum oscillations, thermoelectric coefficients, and the {F}ermi surface of semimetallic {WT}e$_2$, Physical Review Letters \textbf{114}, 176601 (2015).
\bibitem{kumar2017}
N. Kumar, Y. Sun, N. Xu, K. Manna, M. Yao, V. S{\"u}ss, I.
Leermakers, O. Young, T. F{\"o}rster, M. Schmidt, H. Borrmann,
B. Yan, U. Zeitler, M. Shi, C. Felser, and C. Shekhar, Extremely high magnetoresistance and conductivity in the type-{II} {W}eyl semimetals {WP}$_2$ and {MoP}$_2$, Nature Communications \textbf{8}, 1642 (2017).
\bibitem{schonemann2017}
R.~Sch\"onemann, N.~Aryal, Q.~Zhou, Y.-C. Chiu, K.-W. Chen, T.~J. Martin, G.~T. McCandless, J.~Y. Chan, E.~Manousakis, and L.~Balicas, Fermi surface of the {W}eyl type-{II} metallic candidate {WP}$_2$, Physical Review B \textbf{96}, 121108 (2017).
\bibitem{collignon2019}
C. Collignon, X. Lin, C.~W. Rischau, B. Fauqu{\'e} and K. Behnia, Metallicity and superconductivity in doped strontium titanate, Annual Review of Condensed Matter Physics \textbf{10}, 25-44 (2019).
\bibitem{tsujii2003}
N.~Tsujii, K.~Yoshimura, and K.~Kosuge, Deviation from the {K}adowaki-{W}oods relation in {Y}b-based intermediate-valence systems, Journal of Physics: Condensed Matter \textbf{15}, 1993 (2003).
\bibitem{hatzopoulos1985}
Z.~Hatzopoulos and J.~E. Aubrey, Size effects in the electrical resistivity and mean transverse electric field ratio of bismuth and antimony, Journal of Physics F: Metal Physics \textbf{15}, 1093-1101 (1985).
\bibitem{Liang2015}
T. Liang, Q. Gibson, M.~N. Ali, M. Liu, R.~J. Cava, and N.~P. Ong.
\newblock Ultrahigh mobility and giant magnetoresistance in the Dirac semimetal{C}d$_3${A}s$_2$, Nature Materials \textbf{14}, 280-284 (2015).
\bibitem{heremans1977}
J.~Heremans, J-P. Issi, A. A. M. Rashid, and G. A~Saunders, Electrical and thermal transport properties of arsenic, Journal of Physics C: Solid State Physics \textbf{10}, 4511 (1977).
\bibitem{black1971}
M.~A. Black, H.~E. Hall, and K. Thompson, The viscosity of liquid helium 3, Journal of Physics C: Solid State Physics \textbf{4}, 129-142 (1971).
\bibitem{abel1967}
W.~R. Abel, R.~T. Johnson, J.~C. Wheatley, and W.~Zimmermann,
Thermal conductivity of pure ${\mathrm{He}}^{3}$ and of dilute solutions of ${\mathrm{He}}^{3}$ in ${\mathrm{He}}^{4}$ at low temperatures, Physical Review Letters \textbf{18}, 737-740 (1967).
\bibitem{mcCollum1967}
D.~C. McCollum and W.~A. Taylor, Low-temperature specific heat of antimony, Physical Review \textbf{156}, 782-784 (1967).

\end{thebibliography}
\end{document}